\documentclass[%
 preprint, linenumbers,
 amsmath,amssymb,
 aps, physrev,
]{revtex4-2}

\usepackage{graphicx}
\usepackage{dcolumn}
\usepackage{bm}
\usepackage[perpage]{footmisc}

\usepackage{lineno}

\usepackage{perpage}
\usepackage{marginnote}

\usepackage[utf8]{inputenc}
\usepackage[T1]{fontenc}
\usepackage{mathptmx}
\usepackage{etoolbox}
\usepackage{siunitx}
\usepackage{subcaption}

\begin{document}
\nolinenumbers
\preprint{APS/123-QED}

\title{\textbf{Coalescence of viscoelastic drops on a solid substrate} 
}%

\author{Peyman Rostami}
 \email{rostami@ipfdd.de}
\affiliation{ 
Leibniz-Institut f\"ur Polymerforschung  Dresden e.V., Hohe Str. 6, 01069 Dresden, Germany  
}

\author{Alexander Erb}%
\affiliation{ 
Physics of Surfaces,  Institute of Materials Science, Technical University of Darmstadt, Peter-Grünberg-Str. 16, 64287 Darmstadt, Germany
}%

\author{Reza Azizmalayeri}%
\affiliation{ 
Leibniz-Institut f\"ur Polymerforschung Dresden e.V.,  Hohe Str. 6, 01069 Dresden, Germany
}
\author{Johanna Steinmann}%
\affiliation{ 
Physics of Surfaces, Institute of Materials Science, Technical University of Darmstadt, Peter-Grünberg-Str. 16, 64287 Darmstadt,  Germany
}%

\author{Robert W. Stark}%
\affiliation{ 
Physics of Surfaces, Institute of Materials Science, Technical University of Darmstadt, Peter-Grünberg-Str. 16,   64287 Darmstadt, Germany 
}%

\author{Günter K. Auernhammer }
\email{auernhammer@ipfdd.de}
\affiliation{ Leibniz-Institut  f\"ur Polymerforschung Dresden e.V.,  Hohe Str. 6 ,  01069 Dresden,  Germany 
}

\date{\today}

\begin{abstract}

This study investigated the coalescence of polymer solution drops on the solid substrates.
When two drops meet at their contact line on a substrate, the liquid bridge connecting the two drops increases in size with time.
The height and radius of the liquid bridge have a power law dependence on time. 
In the early stage of drop coalescence, the exponents $\alpha$ and $\beta$ of the power law are influenced by the properties of the polymer solution.
We argued that the balance between capillarity and viscoelasticity controls the process, where viscoelasticity must be considered in its full time and shear-rate dependency. 
As a simple proxy for the processes involved, the ratio of the polymer relaxation time to the viscous time of the drop, the elastocapillary ($\textup{Ec}$), is instructive.
In the vicinity of $\textup{Ec}= 1$ the exponents showed a minimum and increased to lower and higher values of $\textup{Ec}$.
A similar dependency is observed for the damping timescales of the capillary waves.
For high elastocapillary numbers, i.e. high polymer relaxation time, drop coalescence on short timescales behaved as in the low viscosity case.
In addition, the bridge profile is influenced by a combination of factors including surface tension, viscosity and polymer stress. 
In summary, we have shown that drop coalescence is strongly influenced by the viscoelasticity of the drops.

\end{abstract}

\maketitle

\section{Introduction}

\begin{figure*}
\centering
\includegraphics[width=0.5\linewidth]{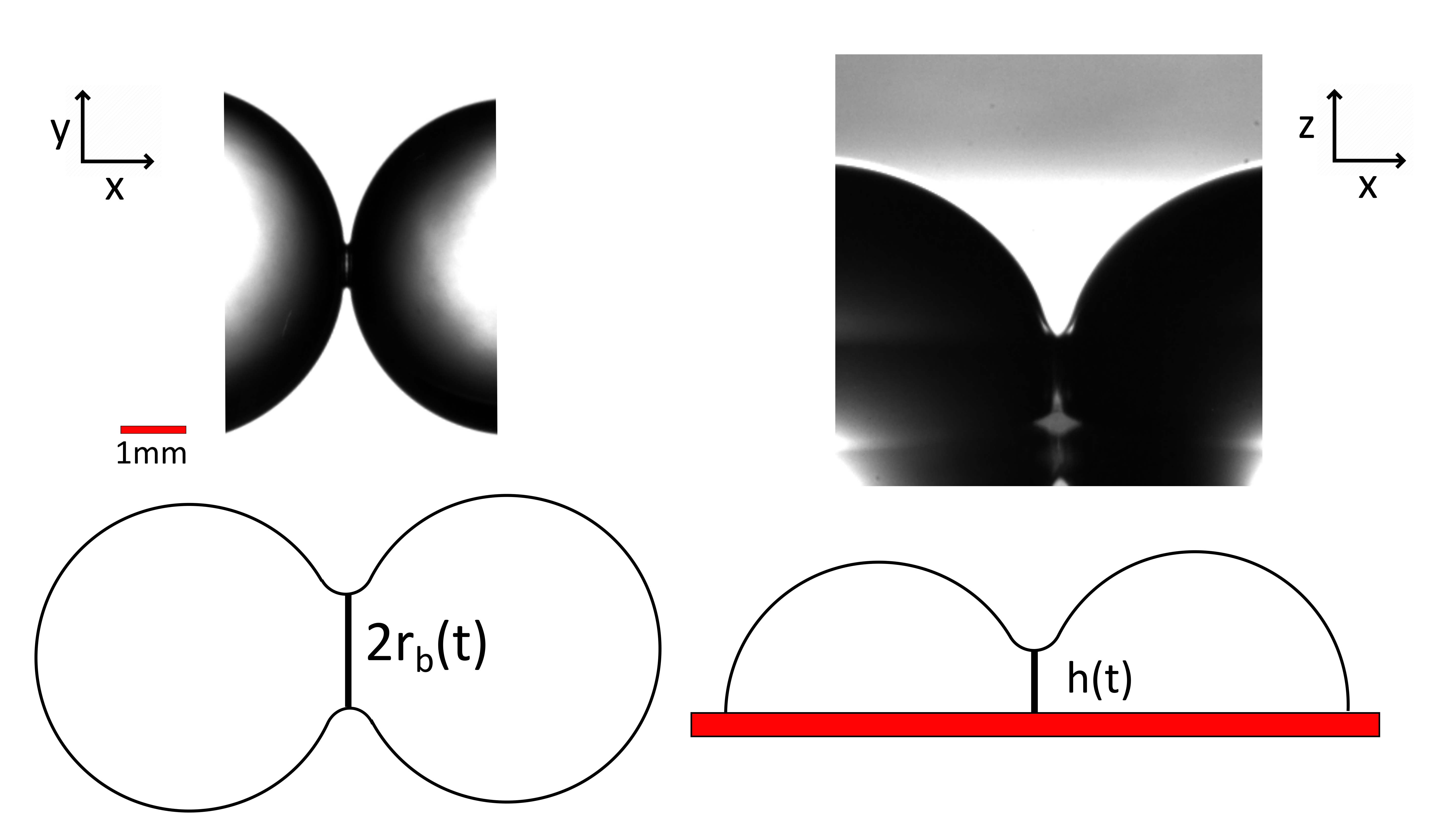}
\caption{ Typical images of drop coalescence and corresponding schematics of the process. Left: top view, right: side view.}
\label{Drop-schematic}
\end{figure*}

The coalescence of drops on a solid substrate plays a crucial role in many industrial applications, from inkjet printing \cite{de2004inkjet,sivasankar2023coalescence} to coating \cite{kamp2017drop}.
Drop coalescence on the substrates involves rapid deformation of the contact lines after the initial moment of contact, followed by the slow relaxation process of the coalesced drop. 
In practice, the coalesced drop may only partially recover a spherical shape due to pinning and contact angle hysteresis of the substrate \cite{zhao1995droplet}.
After the first moment of contact, a contact point forms, leading to the formation of a liquid bridge with radius $r_{b}$ and height $h$, Fig.~\ref{Drop-schematic}.
For simple liquids, the dynamics of the neck radius and height depend on the dominant forces and therefore on the physical properties of the drops (surface tension $\sigma$, viscosity $\eta$ and density $\rho$) as well as on the wettability of the substrate.
For free-hanging drops, the neck dynamics is a function of the square root of time $r_{b}(t)\sim t^{0.5}$ for the normally observed inviscid case (inertia dominated regime) \cite{eggers1999coalescence} and $r_{b}(t)\sim t$ for viscous case \cite{paulsen2011viscous} at very short times.
A transition between these two regimes depends on the fluid properties and the drop size \cite{paulsen2011viscous}.
This transition time is described by the Reynolds number $\textup{Re}\sim\frac{\rho u r_{b}}{\eta } $, where $u$ is the spreading velocity.
Also in the viscous dominant regime, the velocity scales with $u \sim \frac{\sigma}{\eta } $, resulting in a crossover time of $t_{\eta}=\frac{\eta ^{3}}{\rho \sigma^{2}}$.
This time scale for water drop is in order of \SI{0.1}{\nano \second} \cite{aarts2005hydrodynamics}. 
There is no optical way to capture this transition with high speed imaging techniques.

In the case of sessile drop coalescence, the liquid bridge radius (from top view $r_{b}$) and the height (from side view $h$) have different dynamics. 
For the two identical Newtonian drops, the bridge radius grows by the square root of time ($r_{b}\sim t^{0.5}$) \cite{ristenpart2006coalescence,narhe2008dynamic}. 
Other groups reported smaller exponents for the  liquid bridge radius ($\alpha=0.25$) \cite{lee2012coalescence}.
The dynamics of the liquid bridge height is completely different and depends on the drop shape and the symmetry or asymmetry of two drops \cite{eddi2013influence,hernandez2012symmetric}. 
Eddi et al. \cite{eddi2013influence} showed that for two symmetrical drops of low viscosity fluids, the coalescence dynamics depends on the drop contact angle (i.e. surface energy of the substrate). 
For the contact angles lower than \SI{90}{\degree}, $h(t) \sim t^{\frac{2}{3}}$ and the dependence transients to $h(t) \sim t^{0.5}$ for contact angles of the order of \SI{90}{\degree} and higher. 

In most applications, such as 3D printing technology, the working fluids are non-Newtonian \cite{mckinley2005visco, tekin2008inkjet}, i.e., their rheological properties (e.g., viscosity) are not constant as a function of shear rate or strain history.  
To describe the rheological behavior of these systems the intrinsic viscosity $[\eta]$ is helpful, which is defined as $[\eta] = \lim_{c\rightarrow 0} (\eta - \eta_s)/(\eta_s c)$, including the solvent viscosity $\eta_s$, the solution viscosity $\eta$ and the concentration $c$. 
Note that the intrinsic viscosity has the inverse units of a concentration. 
The inverse of this intrinsic viscosity is often called the critical concentration ($c^{*}=\frac{1}{[\eta]}$). 
The entanglement concentration is estimated to $c_{e}=6c^{*}$, so all of our samples were in the semi-dilute range \cite{arnolds2010capillary}, i.e., only slightly entangled. 
The intrinsic viscosity can also be derived from the Mark-Houwink-Sakurada correlation as a function of molar mass  ($M_{w}$) \cite{brandrup1999polymer,tirtaatmadja2006drop}. 
\begin{equation}
[\eta]=0.072M_{w}^{0.65}
\label{MHS-Eq}
\end{equation}

The coalescence of two Newtonian drops can be described by the balance between the capillary force, inertia and the viscous dissipation.
Therefore, the intrinsic time scale of drop coalescence is $t_{vc}= \frac{\eta d}{\sigma}$. 
This time scale is known as the viscous-capillary time scale, where $\eta$ is the viscosity, $\sigma$ is the surface tension and $d$ is the characteristic length scale.
When a deformation is applied to a viscoelastic fluid, the elastic response occurs immediately and only after a certain time (e.g. polymer relaxation time) the viscous part becomes important \cite{ferry1980viscoelastic, strobl1997physics,snoeijer2020relationship}. 
This concept is implemented in various models (e.g., the Maxwell fluid model \cite{oldroyd1950formulation,bird1987dynamics} or the Kelvin-Voigt model \cite{karato2010rheology} for viscoelastic materials). 
The polymer relaxation time is an intrinsic time scale of the fluid.
The elastocapillary number $\textup{Ec}$ relates these two time scales and is defined as the ratio of the polymer relaxation time scale (internal time scale) to the visco-capillary time scale (experimental time scale), Eq.~\ref{Ec}.
For low $\textup{Ec}$, the system is viscosity dominated, for high $\textup{Ec}$, the elasticity dominates.
In between of these two extremes both viscosity and elasticity are dominant.  

\begin{equation}\label{Ec}
\mathrm{Ec}=\frac{\sigma \tau_{ve} }{\eta d}
\end{equation}

One way of interpreting drop coalescence dynamics is to compare it to drop spreading.
This comparison is initiated by the assumption that when two drops of the same fluid collide, a plane with stagnation flow is formed in the coalescence plane.
In this model it is assumed that this plane can behave like the solid substrate in the case of drop spreading \cite{eggers1999coalescence,sivasankar2021coalescence}.
We have previously shown that the drop spreading exponent decreases with increasing elastocapillary number up to $\textup{Ec}\approx 1$. 
After reaching a minimum around $\textup{Ec}\approx 1$, the exponent increases rapidly and reaches a plateau for high elastocapillary numbers \cite{rostami2024spreading}.

Only recently a few studies have been carried out on sessile drop coalescence of non-Newtonian liquids. 
Kern et. al. \cite{kern2022viscoplastic} studied the effect of viscoplasticity on the drop coalescence on the solid substrate.
They showed that the liquid bridge height evolves like Newtonian liquids until it stops at a certain threshold.
This threshold is a function of the fluid's yield stress ($\tau_{y}$), the surface tension ($\sigma$) and the drop contact angle.
A few studies on the viscoelastic drops coalescence have been carried out recently \cite{varma2021coalescence,dekker2022elasticity,varma2022rheocoalescence,varma2022elasticity}, but the results are contradictory. 
Varma et. al. \cite{varma2022rheocoalescence} measured the evolution of liquid bridges connecting a pendant drop to a sessile drop.
They showed that by increasing the polymer concentration up to 20 times of the critical concentration, the coalescence exponent remains constant ($r(t)\sim t^{\alpha}$, $\alpha =0.36$), and then by adding more polymer ($c> 20 c^{*}$), the exponent begins to decrease monotonically.  
From the drop coalescence dynamics, they calculated the relaxation time of the polymer.
The calculated relaxation times are in agreement with the relaxation time derived from rheological measurements as well as from the Zimm model \cite{bird1987dynamics}.  
In an earlier publication by Varma et. al. \cite{varma2021coalescence} correlated the spreading with the concentration ratio to the critical concentration. 
The authors \cite{varma2021coalescence} claimed that for low concentrations ($c/c^{*}< 1$), the exponent of the temporal evolution of the bridge height is $\frac{2}{3}$ and by increasing the polymer, the exponent decreases and reaches $\frac{1}{2}$.
Another group (Dekker et al. \cite{dekker2022elasticity}), claimed that the presence of polymers does not change the temporal evolution of the drop coalescence and it only changes the spatial evolution.
Later, Varma et. al. \cite{varma2022elasticity} (the same author as the previously mentioned publication \cite{varma2021coalescence}), tried to explain the difference between the two publications by attributing the difference to the drop shape before performing the experiment.
Although we cannot fully confirm this, since in the work of Dekker et. al. \cite{dekker2022elasticity}, they stated that the drop deformation due to the needle is local.

These contradictions in the state of the art motivated us to study the drop coalescence of two identical viscoelastic drops on a hydrophobic substrate.
Does the drop coalescence of viscoelastic liquids show a similar dependence on the elastocapillary number $\textup{Ec}$ as some of us found in a previous work? \cite{rostami2024spreading}
Since the viscosity of viscoelastic fluids depends on the shear rate, and the liquid inside the bridge experiences different shear rates during drop coalescence, we studied the dynamics of the liquid bridge in both top and side views.
Furthermore, we studied the effect of molar mass and polymer concentration on the drop coalescence process.
We also investigated the effect of viscoelasticity on the capillary wave during the drop coalescence and investigated the effect of waiting time before the drop coalescence.

\section{\label{sec:level1}Materials and Methods:}

We recorded the drop coalescence process from two orthogonal directions (top and side view) using two high-speed cameras (FASTCAM Mini AX 100, Photron) and illumination systems (consisting of LED lamp (SCHOTT KL 2500) and diffuser sheet). 
The cameras are equipped with Navitar lenses with up to $12\times$ magnification and a $2\times$ F-mount adapter.
All the experiments are recorded at 10,000 frames per second.
The schematic of the drop coalescence from side and top view and the camera output are shown in Fig.~\ref{Drop-schematic}. 
The experiments are performed in a lab with controlled environmental conditions: relative humidity of $48\,\pm\,3\%$ and  temperature of $23.0\,\pm\,0.5\mathrm{^{\circ}C}$).

All the experiments are performed on hydrophobized glass substrates (i.e., the contact angle of seated water drops on these substrates is approximately \SI{90}{\degree}). 
More precisely, the receding and advancing contact angles of water drops on these hydrophobized substrates are $\theta_{R} \approx$ \SI{80}{\degree} and $\theta_{A} \approx$ \SI{105}{\degree}. 
One of the critical points in drop coalescence and drop spreading is the definition of the exact contact time as $t = \SI{0}{\second}$ \cite{rostami2024spreading}.
In current work, such as the previous publication  \cite{rostami2024spreading}, we omitted the first four data points. 
This is crucial because in this time frame the drop coalescence occurs on the solid substrate, not in the free air, i.e., the two drops share a common contact line on the substrate. 
To prepare these substrates, the glass slides (Menzel-Gläser, Microscope Slides, 76 × 26 \si{\milli \meter}) are immersed in the following solvents and incubated in an ultrasonic bath for 15 minutes:
Tetrahydrofuran (Acros Organics Co., 99.6\%), acetone (Fisher Scientific Co., 95\%) and isopropanol (Fisher Scientific Co., 95\%).
The cleaned glass substrates placed in a plasma cleaning device (PlasmaFlecto 10, Plasma Technology GmbH) for 10 minutes.
Immediately after the plasma activation process, the substrates incubated with $30\mu L$ of 1H, 1H, 2H, 2H-perfluorooctyltriethoxysilane (Sigma- Aldrich Co. 97\%) in a closed desiccator and placed in an oven at \SI{100}{\celsius} for 9 hours (see our previous publication for more details \cite{rostami2023dynamic}).

\begin{figure}
\includegraphics[width=0.75\linewidth]{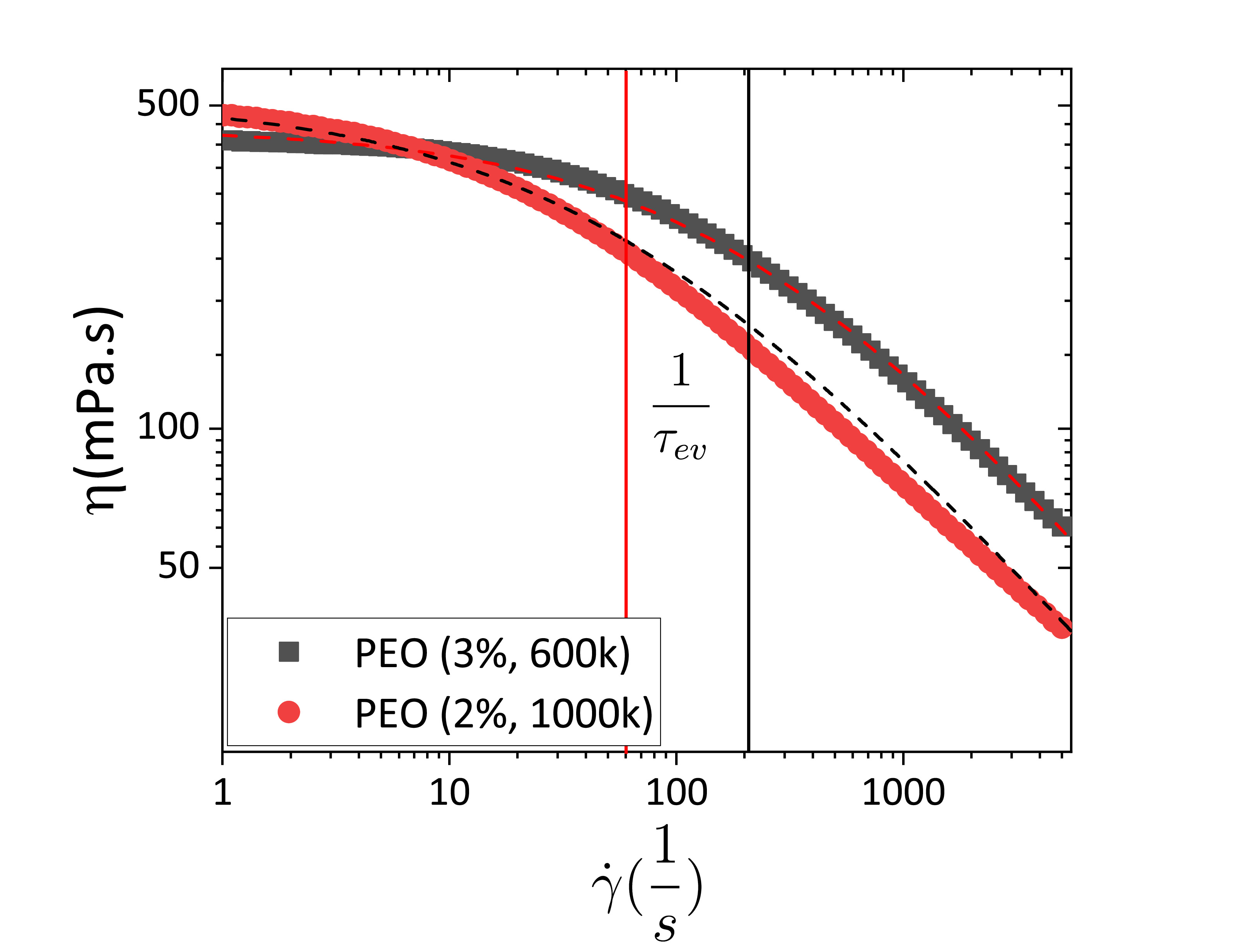}
\caption{\label{Cross-model-graph}The flow curve of different samples of PEO solutions ($\SI{3} {wt\%}$  $\SI{600}{\kilo \gram \per \mol}$ and $\SI{2} {wt\%}$  $\SI{1000}{\kilo \gram \per \mol}$. The vertical lines show the inverse of polymer relaxation time ($\frac{1}{\tau _{ev}}$) as fitted by the Cross model. }
\end{figure}

The preparation of polymer solutions is a time-consuming process \cite{rostami2024spreading}.
In the current study we prepared different samples with different molar masses (300, 600, 1000 and \SI{8000}{(\kilo \gram \per \mol)} and different concentrations.
All polymers were purchased in powder form (Sigma Aldrich Co.) and dissolved in ultrapure water (MicroPure UV/UF, Thermo Scientific Co.).
After ensuring that all the polymers were dissolved, we measured the flow curve of the samples using a commercial rheometer (MCR 502, Anton-Paar GmbH), the flow curve for two samples are ploted in Fig.~\ref{Cross-model-graph}.
Aqueous PEO solutions exhibit a shear thinning behavior, consisting of a Newtonian plateau at low shear rates followed by a shear thinning regime. 
This behavior can be described by standard polymer physics models. 
One of the simplest models was introduced by Cross and is known as the Cross model \cite{cross1965rheology} (Eq.~\ref{cross-fluid-model}).
\begin{equation}\label{cross-fluid-model}
\eta =\frac{\eta _{0}-\eta _{\infty }}{1+(\tau_{ve} \dot{\gamma })^{m}}+\eta _{\infty }
\end{equation}
This model is a four parameter fit: 
The zero shear rate viscosity ($\eta_{0}$), which corresponds to the viscosity at low shear rates (i.e., the Newtonian plateau region), the infinite shear rate viscosity ($\eta_{\infty}$, i.e. viscosity in an assumed plateau at high shear rates), the polymer relaxation time ($\tau_{ve}$, i.e. the time that separates the Newtonian plateau from the shear thinning regime), the rheological exponent ($m$) which determines the slope of the shear thinning part.
This model fits reasonably well our rheological data, Fig.~\ref{Cross-model-graph}. 
The zero shear  viscosity and polymer relaxation time of the samples used are listed in table ~\ref{Physical-Properites}.  
The error on $\tau_{ve}$ is less than 2\% for all given values.

\begin{table}
\caption{\label{Physical-Properites}Composition of operating fluids, the molar masses (as given by the supplier), the zero-shear viscosity $\eta_{0}$  (at \SI{23}{\celsius} and $\dot{\gamma }= 1 (\frac{1}{s})$), the polymer relaxation time $\tau_{ve}$ (s) based on Cross-model (Eq.~\ref{cross-fluid-model}).
}
\centering
\begin{tabular}{lcll}
\hline
\hline
\begin{tabular}[c]{@{}l@{}}Polymer concentration \\ in water (\%)\end{tabular}  & \hspace{1.5cm} \begin{tabular}[c]{@{}l@{}}Molar mass \\ ($\si{\kilo \gram \per \mol}$)\end{tabular} & \hspace{1.5cm} \begin{tabular}[c]{@{}l@{}}$\eta_{0}$ \\ (\SI{}{\milli\pascal\cdot\second}))\end{tabular}     & \hspace{1.5cm} $\tau _{ve}$ (s)    \\
\hline
 PEO ($3\%, 300k$) &\hspace{1.5cm} $300$  & \hspace{1.5cm} $101\pm 0.5$  & \hspace{1.5cm} 0.0002186   \\
 PEO ($3\%, 600k$) &\hspace{1.5cm} $600$  &\hspace{1.5cm} $537\pm 0.5$   & \hspace{1.5cm}0.004355    \\
 PEO ($4\%, 600k$) &\hspace{1.5cm} $600$  & \hspace{1.5cm} $1324\pm 0.5$  & \hspace{1.5cm} 0.02218   \\
 PEO ($1\%, 1000k$) &\hspace{1.5cm} $1000$  & \hspace{1.5cm} $54\pm 0.5$  & \hspace{1.5cm} 0.002646  \\
 PEO ($2\%, 1000k$) &\hspace{1.5cm} $1000$  & \hspace{1.5cm} $495\pm 0.5$  & \hspace{1.5cm} 0.0162  \\
 PEO ($3\%, 1000k$) &\hspace{1.5cm} $1000$  & \hspace{1.5cm} $3340\pm 0.5$  & \hspace{1.5cm} 0.0162   \\
 PEO ($0.25\%, 8000k$) &\hspace{1.5cm} $8000$  & \hspace{1.5cm} $210\pm 0.5$  & \hspace{1.5cm} 0.288  \\
 PEO ($0.5\%, 8000k$) &\hspace{1.5cm} $8000$  &\hspace{1.5cm}  $2450\pm 0.5$  & \hspace{1.5cm} 2.31  \\
\hline
\hline
\end{tabular}
 \end{table}

To determine PEO surface activity, we used a confocal Raman microscope (alpha 300R, Witec GmbH, Ulm, Germany). A \qty{523}{\nano\meter}  Nd:YAG laser at \qty{18}{\milli\watt} focused with a Nikon $20\times$ (numerical aperture 0.40) objective was employed. 
The optical resolution was approximately \qty{20+-2}{\mu\meter} in z-direction and \qty{0.8}{\mu\meter} in xy-direction. The spectra were recorded with a integration time of \qty{0.16}{\second} on a spectrometer with \qty{25}{\mu\meter} pinhole, 600 grooves per \si{\milli\meter}, and a blaze wavelength \qty{500}{\nano\meter}.The confocal measurement spot was automatically adjusted first in the x-direction and then in the z-direction (raster scanning). From the multiple spectra recorded at each lateral position, $z$, a mean spectrum was calculated. Afterward, spectra were smoothed using a moving average filter of size 10. In this way, a highly resolved profile of spectral information across the height of the drop was extracted. \qty{10}{\micro\liter} droplets were used. The measuring position was shifted approximately a quarter drop radius from the apex. 

\section{Results and discussion}
\subsection{Viscous drop coalescence}

\begin{figure}
\includegraphics[width=0.75\linewidth]{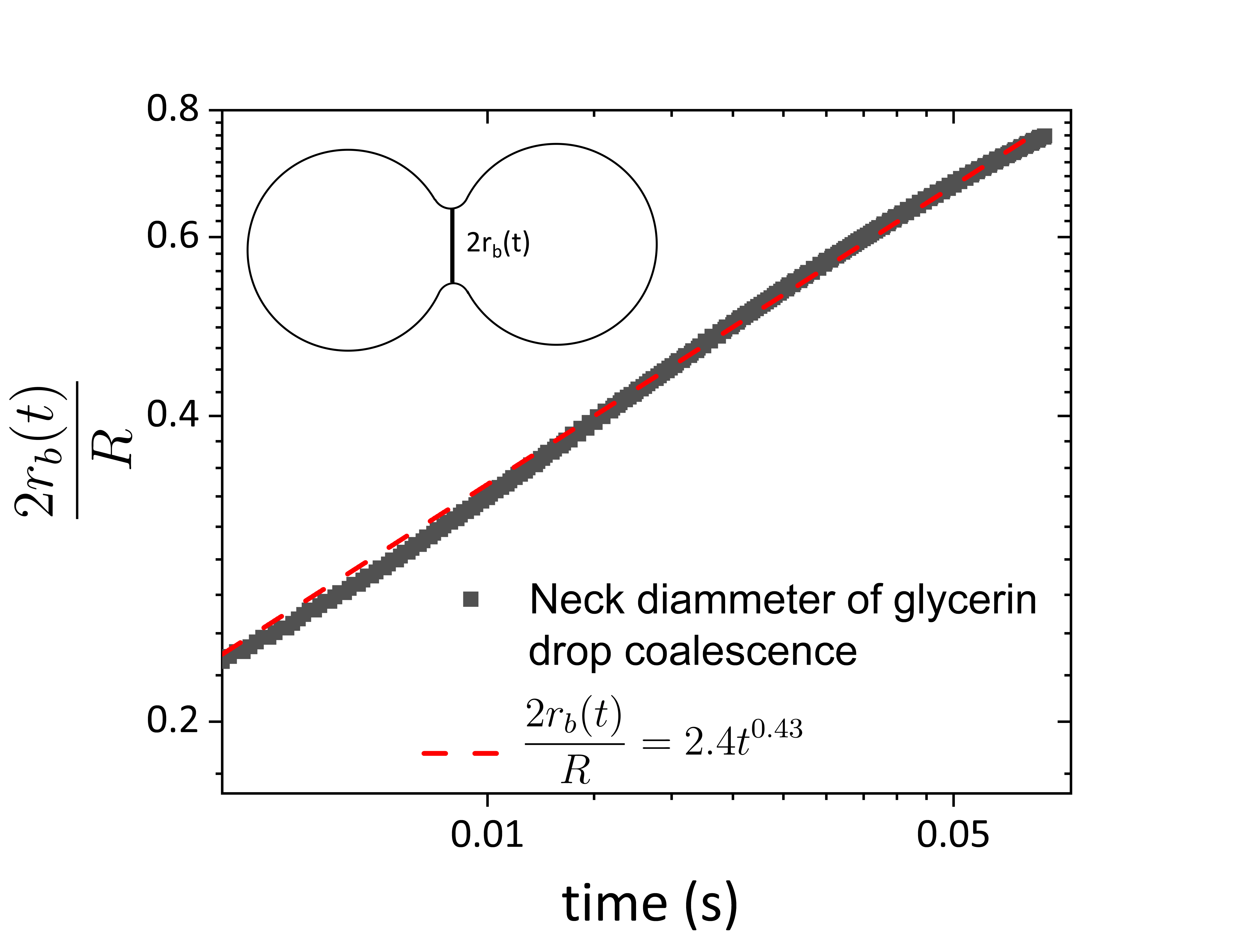}
\caption{\label{Glycerin-Coalescence}Dynamic of the drop coalescence neck radius ($r_b(t)\sim t^{\alpha}$), the insert is the schematic of the drop coalescence from top view.}
\end{figure}

As a reference experiment, we performed the drop coalescence of viscous Newtonian drops (pure glycerine). 
The drops were gently deposited on the substrates and in the late stage of spreading, i.e., at very low velocity, the contact between the drops occurred. 
Due to the large curvature of the liquid surface at the connecting point between two drops, i.e., around the liquid bridge, the capillary pressure forced the liquid into the necking area and led to the growth of the bridge. 
Since the contact angles of the drops were near \SI{90}{\degree}, the liquid bridge width behaved similar the coalescence of pendant drops.
For the Newtonian viscous liquids, the liquid bridge width $r_b(t)$ is expected to be a function of the square root of time ($\sqrt{t}$).\cite{eggers1999coalescence}
We have performed the glycerin drop coalescence experiments on silanized glass substrates, and in general the neck radius dynamics followed ($r_b(t)\sim t^{0.43\pm 0.01}$), which is in a good agreement with the published data.
The results for one of the examples are plotted in Fig.~\ref{Glycerin-Coalescence}. 

\subsection{Viscoelastic drop coalescence}

\begin{table}
\caption{\label{Ec-critical-concentration}The composition of the operating fluids, the sample concentration ($c$) divided by the critical concentration $c^{*}$, the intrinsic viscosity calculated from Eq.~\ref{MHS-Eq} and the elastocapillary number calculated from Eq.~\ref{Ec} by considering the parameters listed in table ~\ref{Physical-Properites}. 
}
\centering
\begin{tabular}{lccc}
\hline
\hline
\begin{tabular}[c]{@{}l@{}}Polymer concentration \\ in water (\%)\end{tabular}  & \hspace{1.5cm}\begin{tabular}[c]{@{}l@{}}$c/c^{*}$\end{tabular}    & \hspace{1.5cm} $[\eta]$ & \hspace{1.5cm} $\textup{Ec}$    \\
\hline
 PEO ($3\%, 300k$) &\hspace{1.5cm} $7.8$  & \hspace{1.5cm} 261.5 & \hspace{1.5cm} $0.07$     \\
 PEO ($3\%, 600k$) &\hspace{1.5cm} $12.3$  & \hspace{1.5cm} 410.3 & \hspace{1.5cm} $0.26$       \\
 PEO ($4\%, 600k$) &\hspace{1.5cm} $16.4$  & \hspace{1.5cm} 410.3 & \hspace{1.5cm} $0.54$    \\
 PEO ($1\%, 1000k$) &\hspace{1.5cm} $5.7$  & \hspace{1.5cm} 571.9 & \hspace{1.5cm} $1.59$    \\
 PEO ($2\%, 1000k$) &\hspace{1.5cm} $11.4$  & \hspace{1.5cm} 571.9 & \hspace{1.5cm} $1.06$    \\
 PEO ($3\%, 1000k$) &\hspace{1.5cm} $17.1$  & \hspace{1.5cm} 571.9 & \hspace{1.5cm} $0.92$     \\
 PEO ($0.25\%, 8000k$) &\hspace{1.5cm} $5.5$  & \hspace{1.5cm} 2209.7 & \hspace{1.5cm} $44.5$    \\
 PEO ($0.5\%, 8000k$) &\hspace{1.5cm} $11$  & \hspace{1.5cm} 2209.7 & \hspace{1.5cm} $30.5$    \\
\hline
\hline
\end{tabular}
 \end{table}

\begin{figure}
   \includegraphics[width=0.75\linewidth]{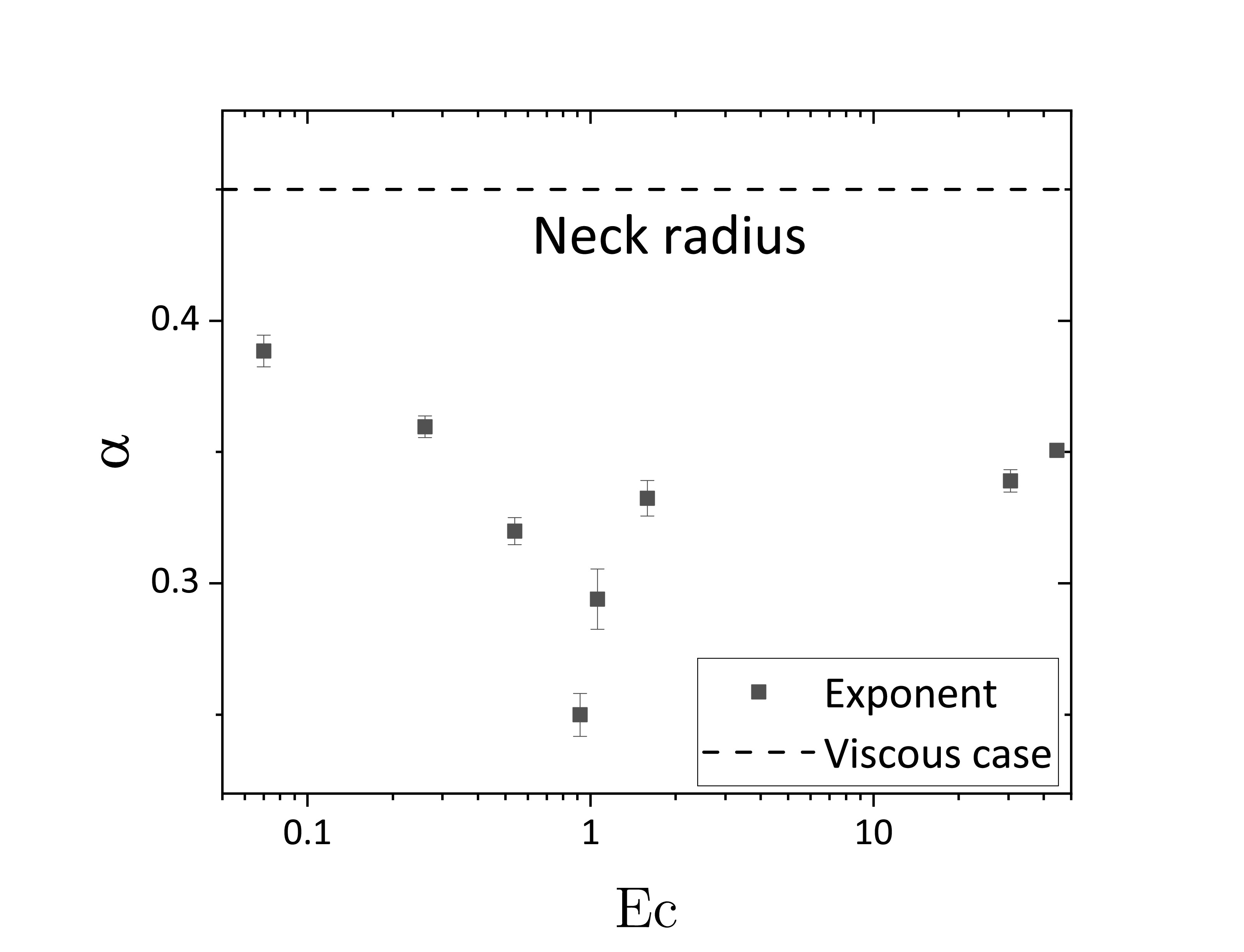}
   \includegraphics[width=0.75\linewidth]{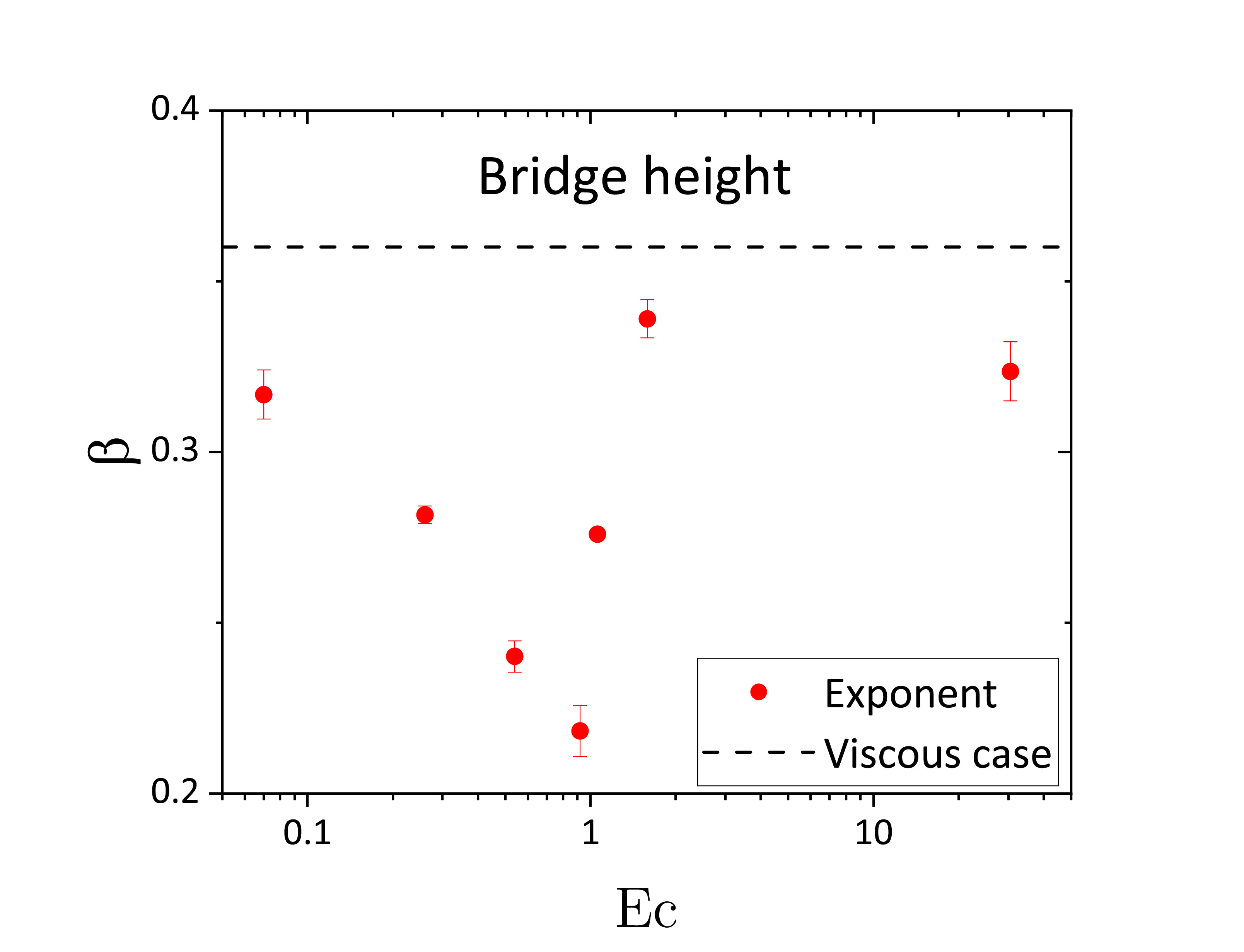}
\caption{Top: The exponent of the liquid bridge neck radius ($\alpha$) as a function of the elastocapillary number \textup{Ec}. 
Bottom: The exponent of liquid bridge height ($\beta$) as a function of the elsatocapillary number (\textup{Ec}). }\label{Exponent-Ec}
\end{figure}

We performed experiments for all liquids listed in the table ~\ref{Physical-Properites} and recorded the process using top view ($r_b(t)\sim t^{\alpha}$) and side view ($h(t)\sim t^{\beta}$) imaging.
After the image processing steps, the results were fitted with a power-law model and the exponents $\alpha$ and $\beta$ were extracted for $r_b(t)$ and $h(t)$, respectively. 
To avoid the effect of zero time definition \cite{rostami2024spreading} as well as artefacts due to capillary waves on the drop surface, we fitted the data in the range of $\SI{0.3}{\milli \second} < t < \SI{20}{\milli \second}$.
The resulting exponents are plotted in Fig.~\ref{Exponent-Ec}. 
The results show that the for the low elastocapillary number (the viscosity dominated regime $\textup{Ec} \ll 1$) and high elastocapillary number (the elasticity dominated regime $\textup{Ec} \gg 1$), the exponents are closer to the Newtonian case. 
However, in the intermediate regime, around $\textup{Ec}  \approx 1$, the both exponents ($\alpha$, $\beta$) decrease and reach a minimum around  $\textup{Ec}  \approx 1$.
The results are in line with what is observed for drop spreading on solid substrates \cite{rostami2024spreading}.
From this result, we concluded that the drop coalescence process and viscoelasticity have the strongest cross coupling when the internal and external characteristic time scales match. 
For low elastocapillary numbers, $\textup{Ec} \ll 1$, the coalescence dynamics is similar to viscous drop coalescence. 
For high elastocapillary numbers, $\textup{Ec} \gg 1$, the polymer solution behaves like a transient hydrogel.

\begin{figure}
\includegraphics[width=0.75\linewidth]{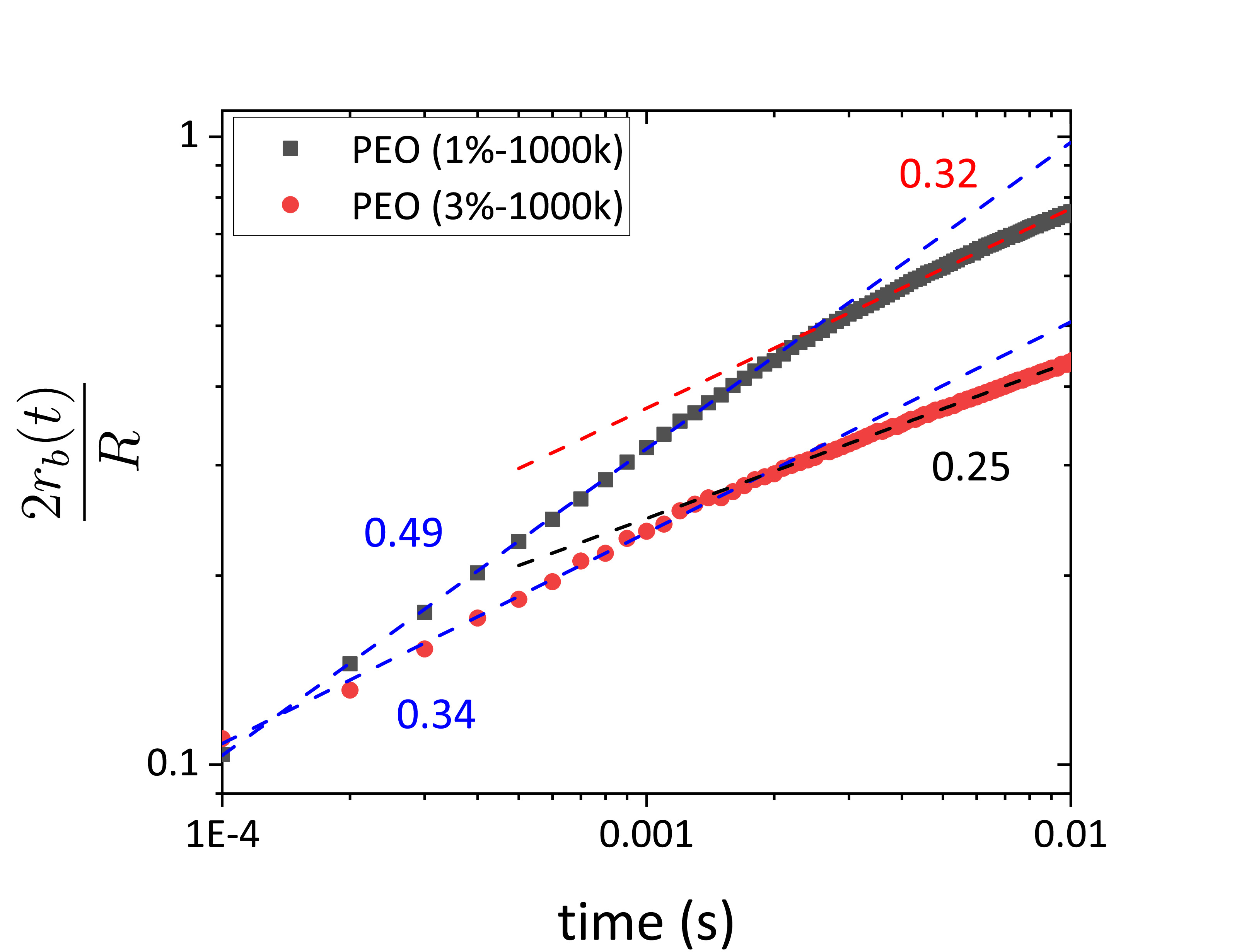}
\caption{\label{second-regime}The normalized neck radius of the liquid bridge divided by the initial drop radius as a function of time. The transition between two regimes is shown for two cases. The exponent of the first regime is higher than that of the second regime, Fig.~\ref{Exponent-Ec} shows the values of the second regime.  }
\end{figure}

At high elastocapillary numbers, i.e., for $M_{w}=8000$ and \SI{1000}{(\kilo \gram \per \mol)}, the exponent of the neck radius ($r_b(t)\sim t^{\alpha}$) differs more from the Newtonian case compared to the liquid bridge height.  
The origin of this difference lies in the appearance of an additional coalescence regime for high $\textup{Ec}$, Fig.~\ref{second-regime}.
In general, the drop coalescence of these high molar mass liquids has an initial regime that spreads faster and with a higher exponent. 
The first regime lasts for a few milliseconds. 
Then, the dynamics slows down and the exponent decreases. 
In Fig.~\ref{Exponent-Ec}, we plotted the exponent values for the second regime in Fig.~\ref{second-regime}.
This phenomenon is also observed for higher molar masses drop spreading \cite{rostami2024spreading}.
This additional regime could be the origin of the difference between our results and those reported by other group \cite{dekker2022elasticity}.
Since their experimental time scale is limited to a few milliseconds, they observed only the first regime and claimed that the addition of polymer does not change the coalescence exponents.
For even  a longer  times as shown in Fig.~\ref{second-regime} the contact line was pinned to the substrate due to high contact angle hysteresis of the substrate. 
The time evolution of a sample (PEO 1\%, 1000 \si{\kilogram \per \mol}) is presented in the SI.

On the short time scales the drop coalescence is a start-up of a deformation process. 
When we start to shear a polymer solution, it is known that it takes some time for the polymer to contribute to the viscosity \cite{ferry1980viscoelastic,costanzo2016shear,VEREROUDAKIS2023105021}. 
At the very beginning of the deformation process, the polymer chains follow the deformation of the solvent. 
On a time scale characterized by the polymer relaxation time, the deformed chains relax and begin to contribute to the viscosity of the polymer solution.  
This effect is also more pronounced at the higher molar masses. 
For this reason, the early stage of drop coalescence ($t <$ \qty{2}{\milli \second}) is a quasi non-viscous process.
This implies that the elastic contribution of the polymers dominates in this early regime. 
The drop behaves like a transient hydrogel in this very early regime.
And only after enough time will the polymers begin to contribute to viscosity and flow.

\subsection{Dynamics of the capillary waves}
\label{sec:DynCapWaves}

\begin{figure}
\includegraphics[width=0.75\linewidth]{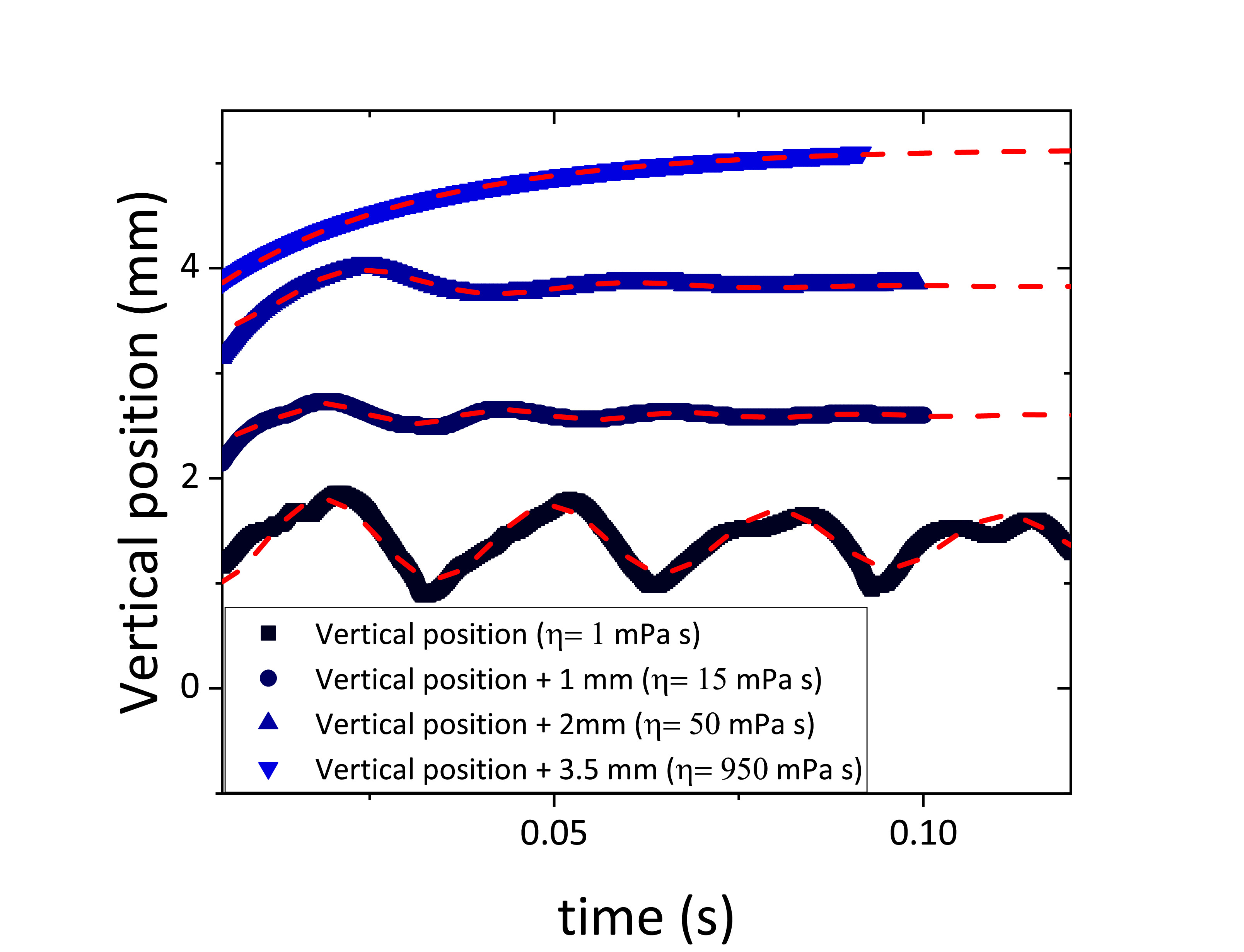}
\caption{\label{Capillary-wave-Newtonian}
The vertical position of the contact point between two drops (measured at the position where the bridge height $h(t)$ was measured in the early stage of coalescence) as a function of time for water and various aqueous glycerin solutions. 
By increasing the viscosity (i.e. adding glycerin to water), the capillary wave damps faster and the wavelength increases.
For better visualization, the neck heights are shifted by the mentioned values and the guide to the eyes indicates the capillary waves.}
\end{figure}

The crossover from the viscosity-dominated regime to the elasticity-dominated regime can also be found in the characteristics of the capillary waves generated by the coalescence process. 
To quantify the capillary waves, we measured the neck height of the bridge between the drops $h(t)$ versus time for longer time scales.
For Newtonian liquids, as the viscosity increases, the amplitude of the capillary waves decreases, the period increases, and the capillary wave dampens faster, see Fig.~\ref{Capillary-wave-Newtonian}. 

In general, the capillary wave has an oscillatory nature, $\sin\left[B(t-t_{c})\right]$, and damps exponentially, $h(t) \approx \sin\left[B(t-t_{c})\right]\exp\left[-\frac{t}{\tau }\right]$, Eq.~\ref{Dampsine}.
\begin{equation}\label{Dampsine}
h(t) =h_{0}+A_h\exp \left( -\frac{t}{\tau } \right) \sin \left[ B(t-t_{c}) \right]
\end{equation}
For the Newtonian cases, the damping timescale ($\tau$) is known to be a function of capillary wave properties such as capillary wave number $\lambda $, viscosity $\eta$, and liquid density $\rho$ \cite{behroozi2003fluid,behroozi2001dispersion} (see Eq.~\ref{Dampingtimescale}).   
\begin{equation}\label{Dampingtimescale}
\tau=\frac{\rho \lambda ^{2}}{8\pi ^{2}\eta }
\end{equation}

\begin{figure}
\includegraphics[width=0.75\linewidth]{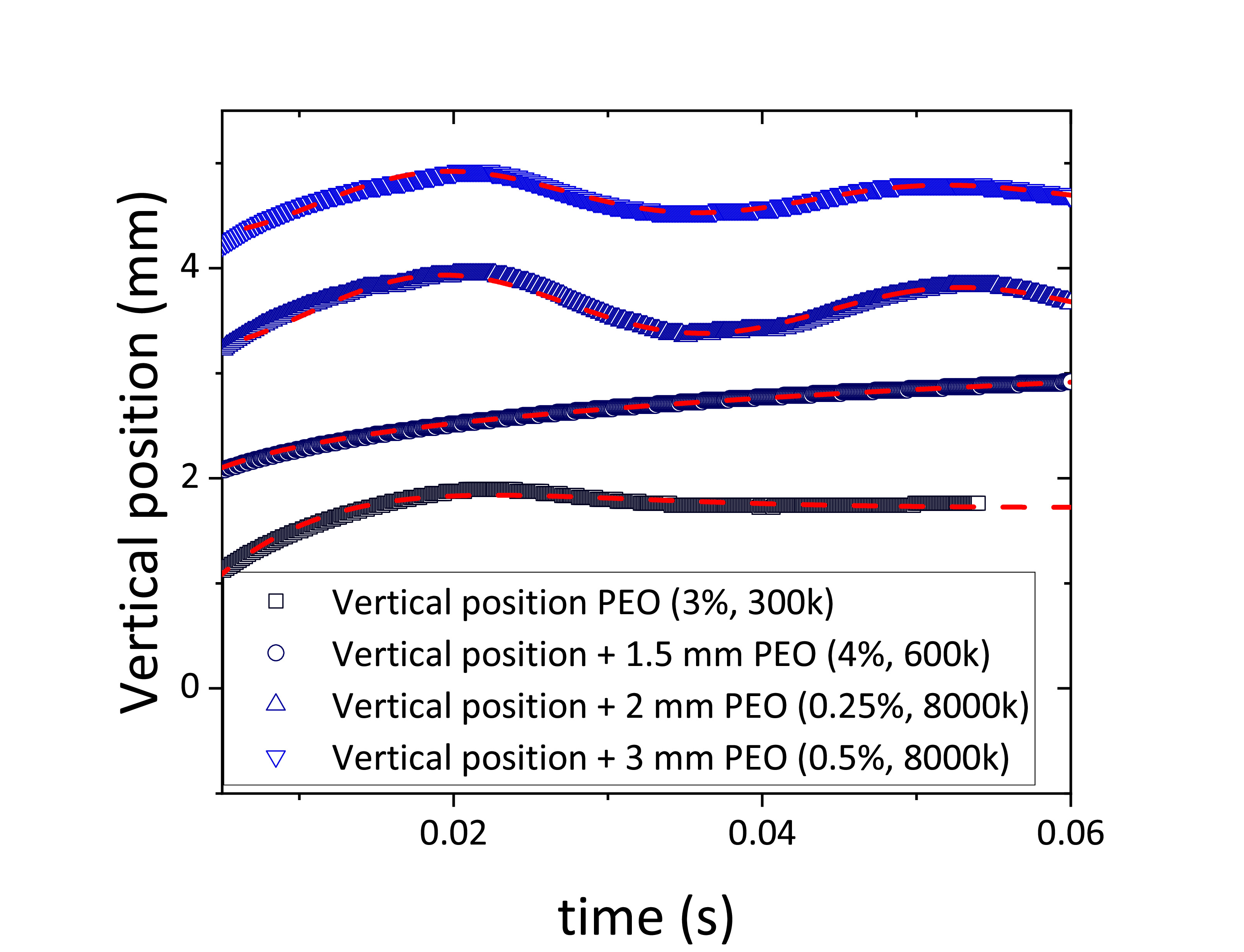}
\caption{\label{Capillary-Wave-Non-Newtonian}
The vertical position of the contact point between two drops (measured at the position where the bridge height $h(t)$ was measured in the early stage of coalescence) as a function of time for different polymer solutions. 
By increasing the viscosity (for intermediate molar masses), the capillary wave damps faster and the wavelength increases.
For high molar masses \SI{8000}{(\kilo \gram \per \mol)}, the capillary waves reappear. 
For better visualization, the exact values are shifted by the mentioned values and the guide to the eyes indicates the capillary waves.}
\end{figure}

The damping timescale ($\tau$) is a function of the liquid viscosity, with increasing viscosity the capillary wave dampens faster, Fig.~\ref{Capillary-wave-Newtonian}. 
In the case of our polymer solutions (for intermediate polymer molar masses $(300 \; \mathrm{to}\; 600)\si{\kilo \gram \per \mol}$, $h(t)$ follows the trends expected for Newtonian liquids. 
As the viscosity $\eta$ increases the damping timescale $\tau$ decreases, Fig.~\ref{Capillary-Wave-Non-Newtonian}. 

\begin{figure}
\includegraphics[width=0.75\linewidth]{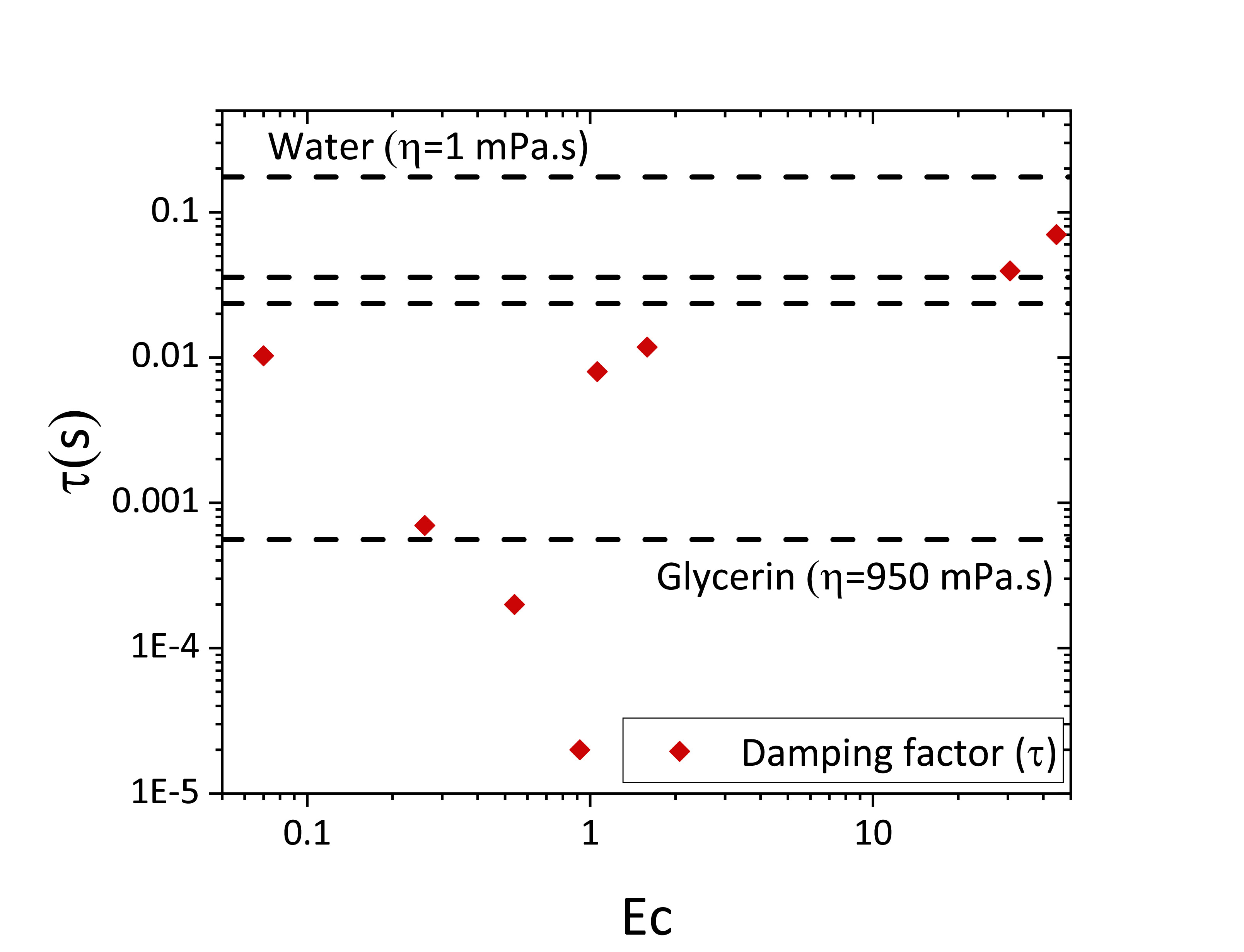}
\caption{\label{Damping-Ec}
The damping timescales $\tau$ of the capillary wave due to the drop coalescence as a  function of the elastocapillary number (\textup{Ec}) for the samples listed in table \ref{Physical-Properites}. 
The values for Newtonian cases are plotted as dashed lines. 
The transition around ($\mathrm{Ec} = 1$) is visible. }
\end{figure}

Based on Eq.~\ref{Dampingtimescale}, the damping timescale should scale with the inverse of the viscosity ($\tau \sim \frac{1}{\eta }$). 
For the Newtonian and the polymer solutions with intermediate polymer chains (i.e. $Ec< 1$), this relation works as predicted (see SI). 
The situation is quite different for samples with higher elastocapillary numbers. 
Although the zero-shear rate viscosity of 0.25\% and 0.5\% of \SI{8000}{(\kilo \gram \per \mol)} is in a range where no capillary waves are expected for Newtonian liquids,  well-developed capillary waves were observed for these polymer solutions. 
This observation is a consequence of the characteristic time scales of the capillary waves and the polymer relaxation. 
In fact, the polymer relaxation time for these two samples (\SI{0.288}{\second} and \SI{2.31}{\second}, respectively, table ~\ref{Physical-Properites}) is well beyond the observation timescale in Fig.~\ref{Capillary-Wave-Non-Newtonian} (about \SI{20}{\milli\second} for one period). 
At the frequency of the capillary waves, the elasticity of the viscoelastic behavior is clearly dominates. 
A collection of different capillary waves for polymer solutions is shown in Fig.~\ref{Capillary-Wave-Non-Newtonian}. 
The damping timescale for all samples listed in table ~\ref{Ec-critical-concentration} is shown in Fig.~\ref{Damping-Ec}. 
The dashed lines in Fig.~\ref{Damping-Ec} indicate the damping timescale for Newtonian cases (aqueous glycerin solutions). 
The last two data points correspond to high molar masses (\SI{8000}{(\kilo \gram \per \mol)}) with zero shear rate viscosities of $210$ and \SI{2500}{\milli \pascal \second}.
The damping factor of these samples is equivalent to the damping factors of Newtonian liquids in the range of \SI{1}{\milli \pascal \second} to \SI{15}{\milli \pascal \second}. 
Thus, in the early stages of drop coalescence, the effective viscosity is less than the zero shear rate viscosity.

\begin{figure}
   \includegraphics[width=0.75\linewidth]{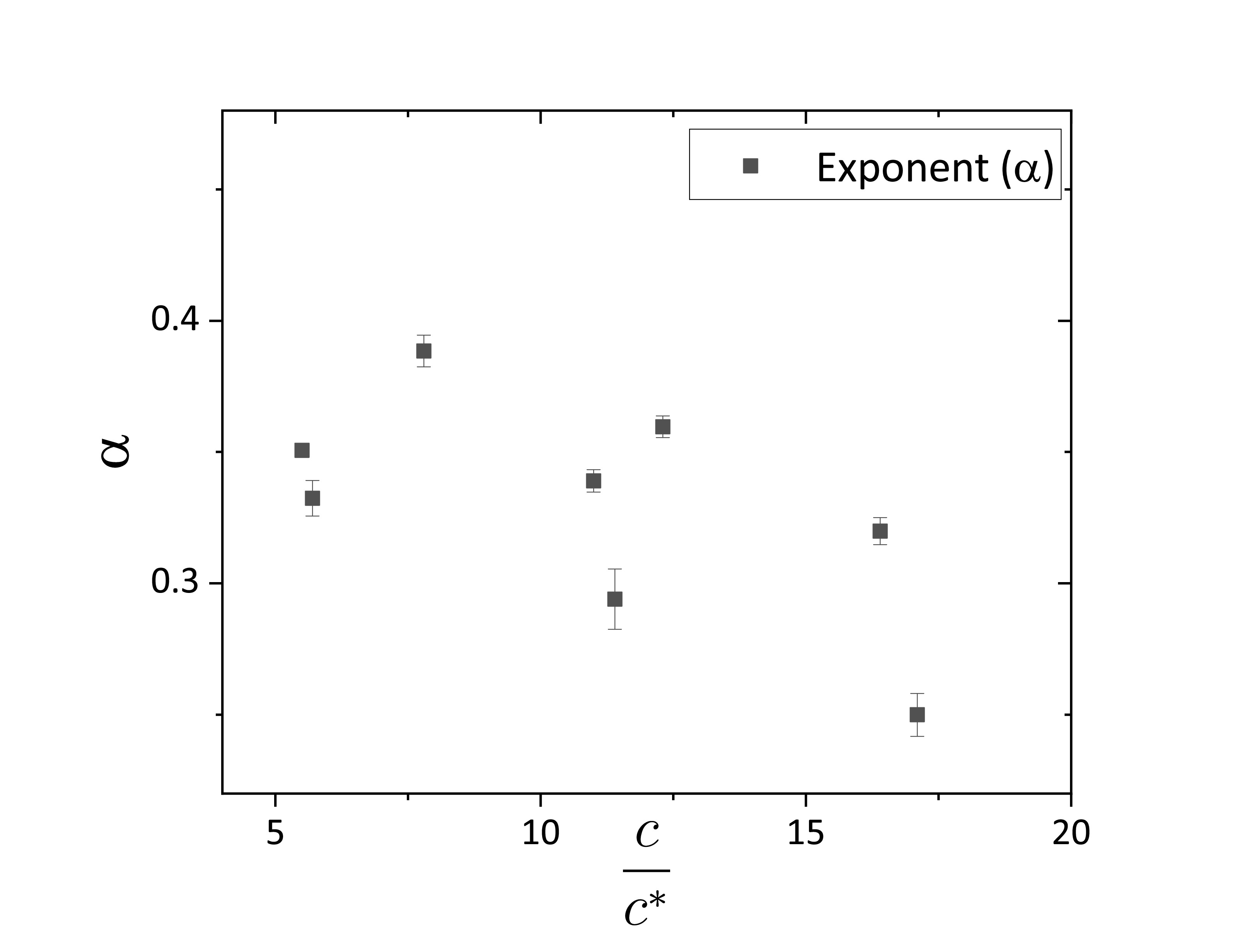}
   \includegraphics[width=0.75\linewidth]{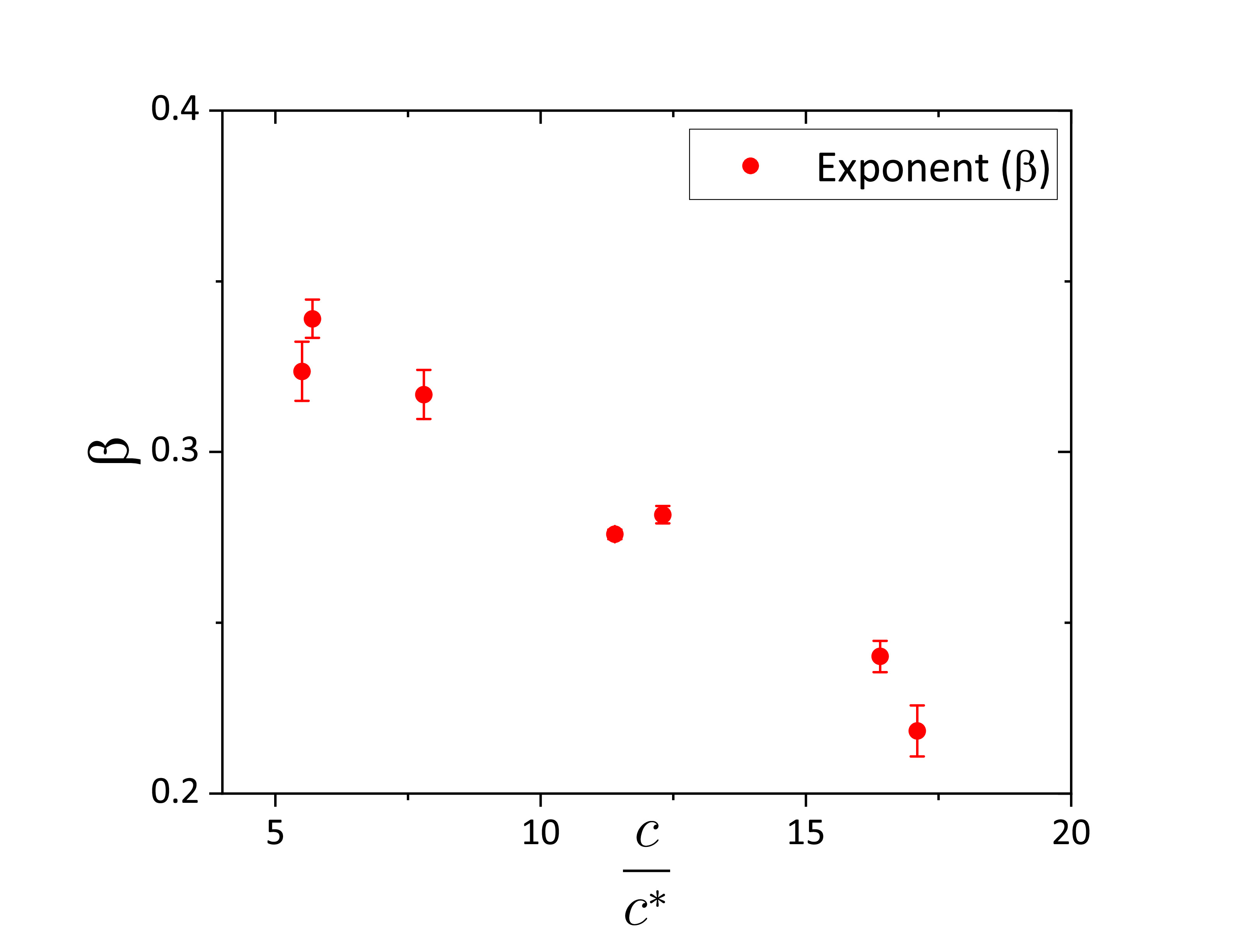}

\caption{Top: The exponent of the liquid bridge neck exponent ($\alpha$) for different polymer concentrations $\frac{c}{c^{*}}$ . 
Bottom: The exponent of liquid bridge height ($\beta$) as a function of the polymer concentration values.}
\label{Exponent-C}
\end{figure}

The other relevant parameter that may be important is the concentration ($\frac{c}{c^{*}}$). 
For the low contact angle substrates, it was shown that the exponent of the liquid bridge height ($\beta$) remains constant and after a threshold ($\frac{c}{c^{*}}> 10$), the exponent decreases monotonically \cite{varma2022elasticity}. 
The same observation applies to the coalescence of a sessile drop and a pendant drop \cite{varma2022rheocoalescence}.
In this configuration, the threshold for the monotonous decrease occurs at higher concentrations \cite{varma2022rheocoalescence} ($\sim \frac{c}{c^{*}}= 20$).
We found the same dependency for the exponent from the side view $\beta$, Fig.~\ref{Exponent-C}. 
As a function of  $\frac{c}{c^{*}}$  we observed a monotonous decrease of $\beta$. 
The threshold in this case is around ($\frac{c}{c^{*}}= 5$) which the difference with the previous cases \cite{varma2022elasticity, varma2022rheocoalescence} can be explained by geometric and configuration differences.
 
For the liquid bridge radius exponent (from the top view, $\alpha$), we could not find a direct dependence on concentration (see Fig.~\ref{Exponent-C}).
The liquid bridge connecting two drops, expands parallel and perpendicular to the substrate.
Parallel to the substrate, the fluid experiences more shear (due to the presence of the solid substrate), which could be the reason for the different observation.
This shows that the elastocapillary number is a better parameter to mimic the drop coalescence dynamics.
The polymer concentration parameter does not take into account the wetting parameters (e.g. drop size and surface tension). 
Also, the polymer concentration does not represent the elasticity of the fluid, but the elastocapillary number combines all of these parameters. 
The elasticity is more important when we have higher shear (in case of liquid bridge neck $r(t)$).

\subsection{PEO surface activity}

To evaluate the PEO surface activity, Raman spectra were recorded along a line from the outside of the droplet into the droplet. The measurement position was offset by approximately a quarter of the drop radius from the center of the droplet. Due to the refraction on the curved surface of the droplet, the actual measured line is angled at approximately \ang{11} relative to the normal. 

\begin{figure}
   \includegraphics[width=0.8\linewidth]{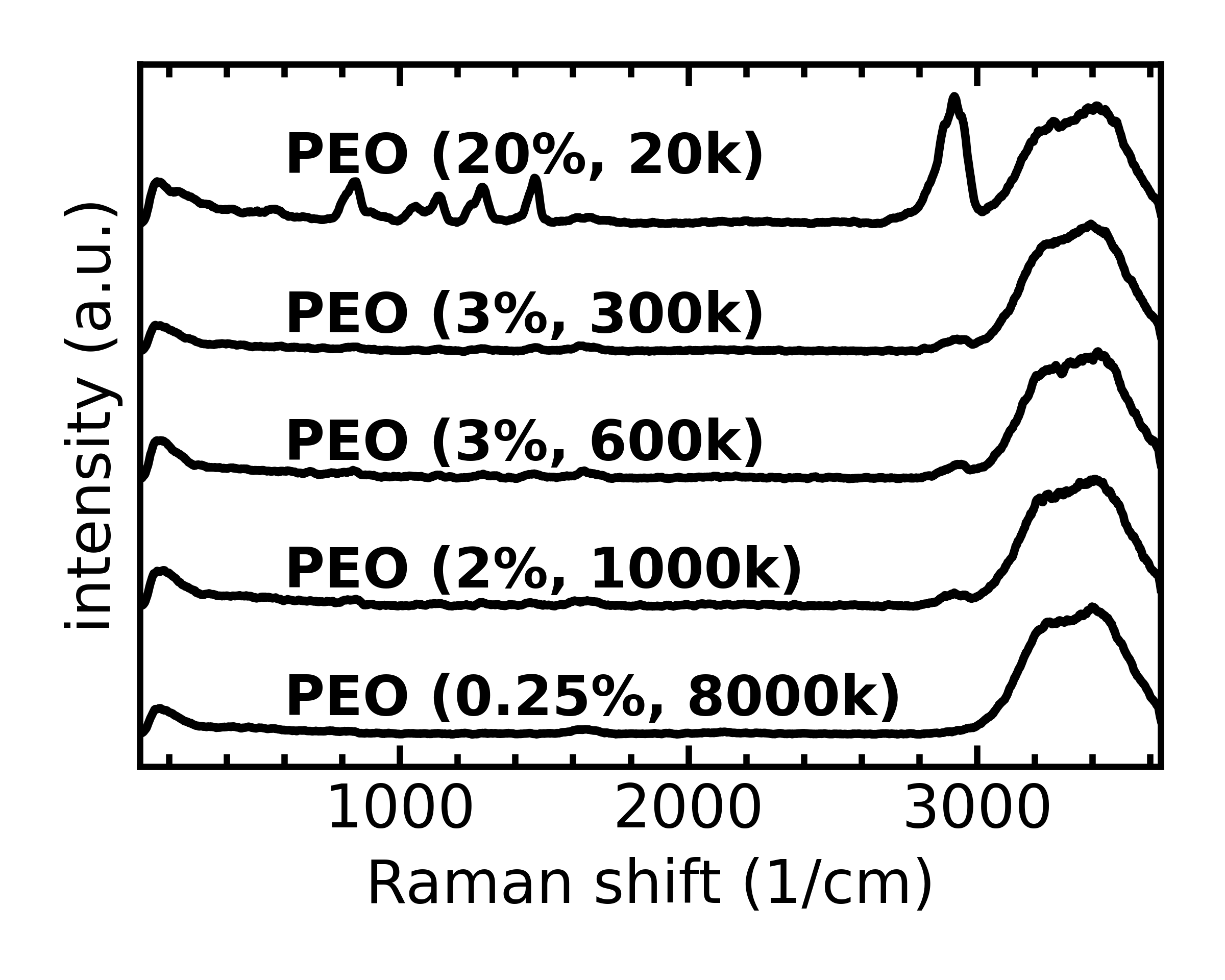}
\caption{Raman spectra of different PEO/water mixtures. The CH-stretching band (approx. \qty[per-mode=power]{2800}{\per \centi \meter}) and OH-stretching band (approx. \qtyrange[per-mode=power]{3000}{3600}{\per \centi \meter}) are visible in all short-chained PEO spectra. With longer PEO chains and thus decreasing absolute concentration of PEO, the intensity of the CH peak decreases compared to the OH peak. Raman spectra were recorded using a 20$\times$/0.40 objective with \qty{17.5}{\milli \watt} laser power and \qty{0.16}{\second} integration time. Spectra were vertically displaced and normalized to their respective OH-band.}
\label{Raman1}
\end{figure}

In order to evaluate the proportion of PEO present in the PEO/water mixtures, the CH-stretching (approx. \qtyrange[per-mode=power]{2800}{3050}{\per \centi \meter}) and the OH-stretching (approx.  \qtyrange[per-mode=power]{3050}{3650}{\per \centi \meter} bands were employed as indicators. Spectra of all PEO/water mixtures used are shown in  Fig.~\ref{Raman1}. For longer-chained PEOs the intensity of the CH band decreases while the OH band intensity remains constant. 



Given that the method requires the presence of both specific bands, the CH band limits its applicability.  As for PEO \qty{8000}{\kilo \gram \per \mol} no CH peak was found, it was not investigated further.  Since PEO \qty{20}{\kilo \gram \per \mol} is spectrally illustrative, it was chosen as a reference for the interpretation of the Raman measurements.

\begin{figure}
\includegraphics[width=0.85\linewidth]{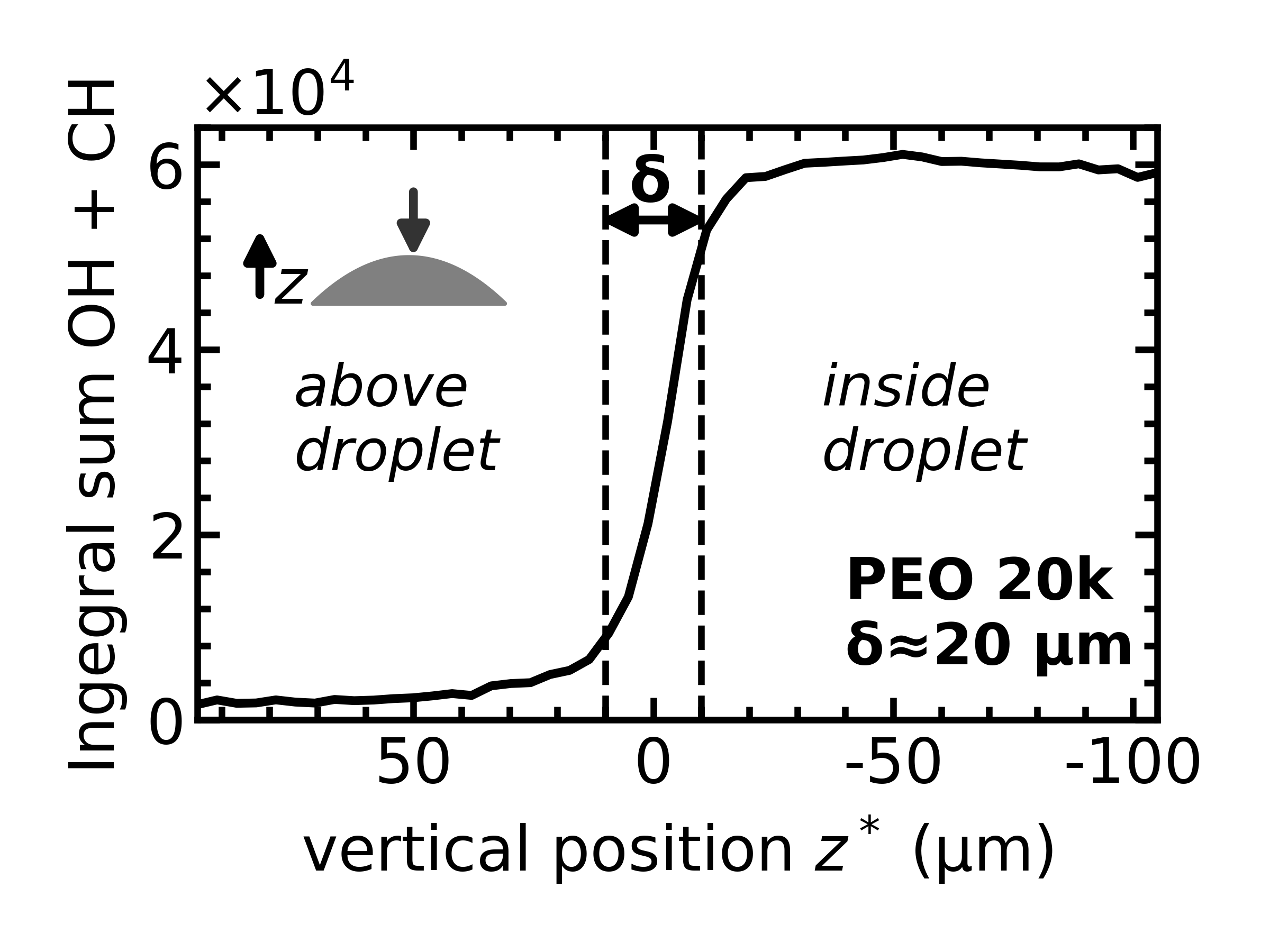}
\caption{Sum of the specific integrals (CH and OH) in the spectral range \qtyrange[per-mode=power]{2800}{3650}{\per \centi \meter} as a function of the vertical sampling position $z^*$. The vertical position $z^*$ was adjusted such that the approximate position of the droplet surface is located at $z^*=0$. The measurement mode was a xz-raster scan, during which 2 measurements were recorded horizontally and 50 measurements vertically in a region of \qtyproduct{4 x 200}{\micro \meter}. The scan position (indicated by a grey arrow in the schematic) was slowly moved from the gas phase into the PEO (20\%, 20k) droplet. The transition width $\delta$ corresponds to the confocal resolution of the device, which was determined as \qty{20+-2}{\micro \meter}. Raman measurements were performed with a 20$\times$/0.40 objective, \qty{17.5}{\milli \watt} and \qty{0.16}{\second} integration time. }
\label{Raman2}
\end{figure}

To determine the lateral resolution, Raman scans along a line from the gas phase into the droplet were performed. Since the Raman response of the atmosphere is negligible in this case, the lateral resolution $\delta$ was determined as the transition width of the count integral sum of both specific bands (CH-stretching and OH-stretching) (Fig. \ref{Raman2}). $\delta$ was determined as \qty{20+-2}{\micro \meter}.

Since no pure substance spectra of the PEOs could be recorded, a quantitative evaluation of the data using spectral deconvolution was impossible. However,  Raman spectra of fluid mixtures (such as aqueous PEO solutions) can be evaluated using the ratio of the specific band integrals as long as the integrals do not overlap (which is not the case in PEO/water mixtures) \cite{numata_quantitative_2011}. The integral ratio correlates with the concentration. Thus, statements can be made about the concentration profiles within the droplet \cite{bell_concentration_2022}. Here, the ratio of the specific bands CH (\qtyrange[per-mode=power]{2800}{3050}{\per \centi \meter}) and OH (\qtyrange[per-mode=power]{3050}{3650}{\per \centi \meter}) was used. 

\begin{figure}
\includegraphics[width=1.05\linewidth]{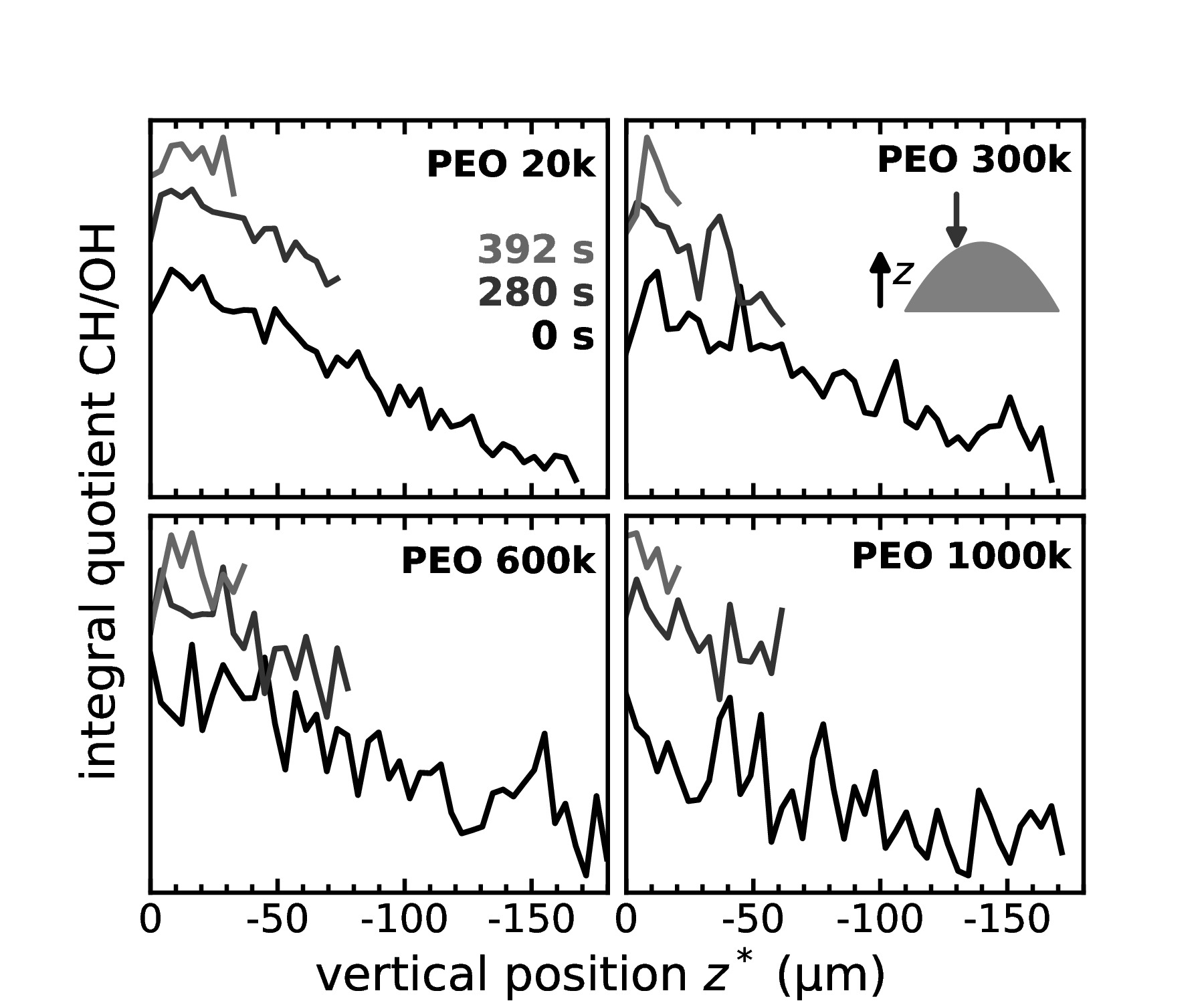}
\caption{Specific Raman band ratio during drop aging for PEO (20\%, 20k), PEO (3\%, 300k), PEO (3\%, 600k), and PEO (2\%, 1000k). The quotient of the specific band integrals CH (\qtyrange[per-mode=power]{2800}{3050}{\per \centi \meter}) and OH (\qtyrange[per-mode=power]{3050}{3650}{\per \centi \meter}) is proportional to the PEO concentration. A higher ratio indicates a higher PEO content. The vertical position $z^*$ was adjusted such that the position of the droplet surface is located at $z^*=0$. The measurement mode was a xz-raster scan, during which 2 measurements were recorded horizontally and 50 measurements vertically in a region of \qtyproduct{4 x 200}{\micro \meter}. The scan position on the droplet is indicated by a grey arrow in the schematic in the upper right plot.  Raman measurements were performed with a 20$\times$/0.40 objective, \qty{17.5}{\milli \watt}, and \qty{0.16}{\second} integration time. 
}
\label{Raman3}
\end{figure}

Raman scans into the droplet (along gravity) were performed. By continuously repeating the measurement, the concentration distribution in the aging droplet was investigated. This was done here with one measurement lasting approximately \qty{56}{\second}. The preparation time to start the first measurement was approximately \qty{15}{\second}. Specific integral ratios for aging droplets of different PEOs as a function of vertical position $z^*$ are shown in Fig.~\ref{Raman3}. $z^*$ was selected such, that the droplet surface is located at approximately $z^*=0$. All aqueous PEO solutions show a decreasing PEO concentration from the droplet surface to the bulk. This behavior has been observed before \cite{bouillant2022rapid, kim_additional_1993, mamalis_effect_2015}.
For aging droplets, the droplet size decreases due to water evaporation, while PEO does not evaporate. This results in shorter curves and higher overall PEO concentrations, which cause the curves to shift upwards. Slightly stronger gradients were found for shorter PEO chain lengths. The noise increases for longer chain lengths, as the CH signal is difficult to detect for long PEO chains (compare Fig.~\ref{Raman1}). 

We found that PEO is surface active and enriches at the droplet surface.  The PEO concentration slowly decreases from the surface towards the droplet bulk. Surface activity of PEO is a known phenomenon. When adding PEO to a water droplet the surface tension decreases \cite{kim1997surface}.  The equilibration of the surface tension to changes in the system is strongly influenced by the diffusion of the PEO molecules in the aqueous medium. The diffusion of PEO molecules to the surface and their transport at the interface is an inverse function of the hydrodynamic radius, $D_{diff} \sim \frac{1}{r_{h}}$.
In addition, the hydrodynamic radius is a function of the molar mass ($r_{h} \sim  M$).
During drop coalescence, the surface area is reduced in the vicinity of the neck. This locally also reduces the surface tension and  imposes a Marangoni stress on the interface.
Since the polymer molecules used here are large, the rate of diffusion and adsorption of polymers into and out of the bulk is low.
The Marangoni stress pulls the liquid away from the liquid bridge connecting the two drops. 
In the next section, we look at the evolution of the bridge profile.

\section{The bridge profile evolution during drop coalescence }

\begin{figure}
   \includegraphics[width=0.75\linewidth]{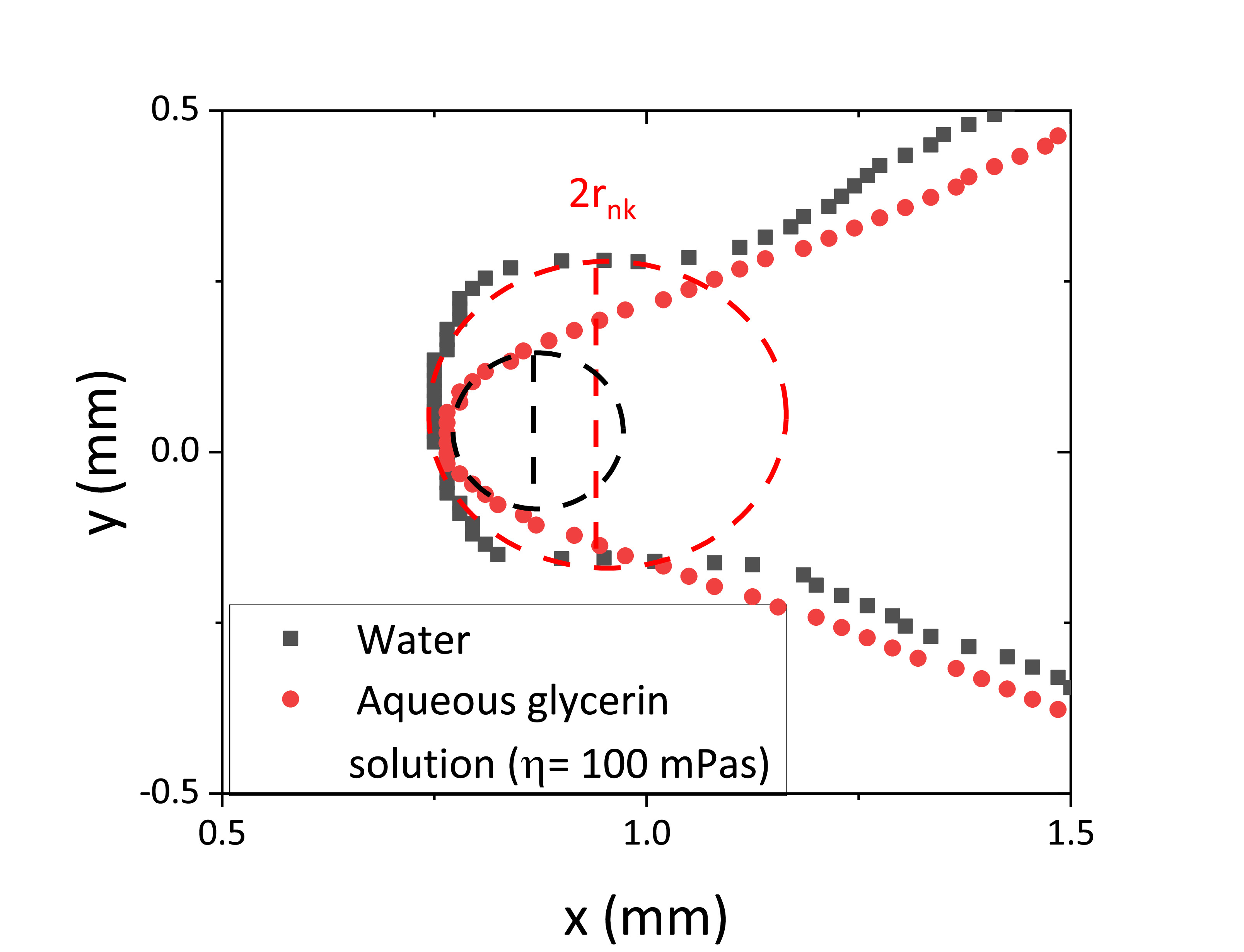}
   \includegraphics[width=0.75\linewidth]{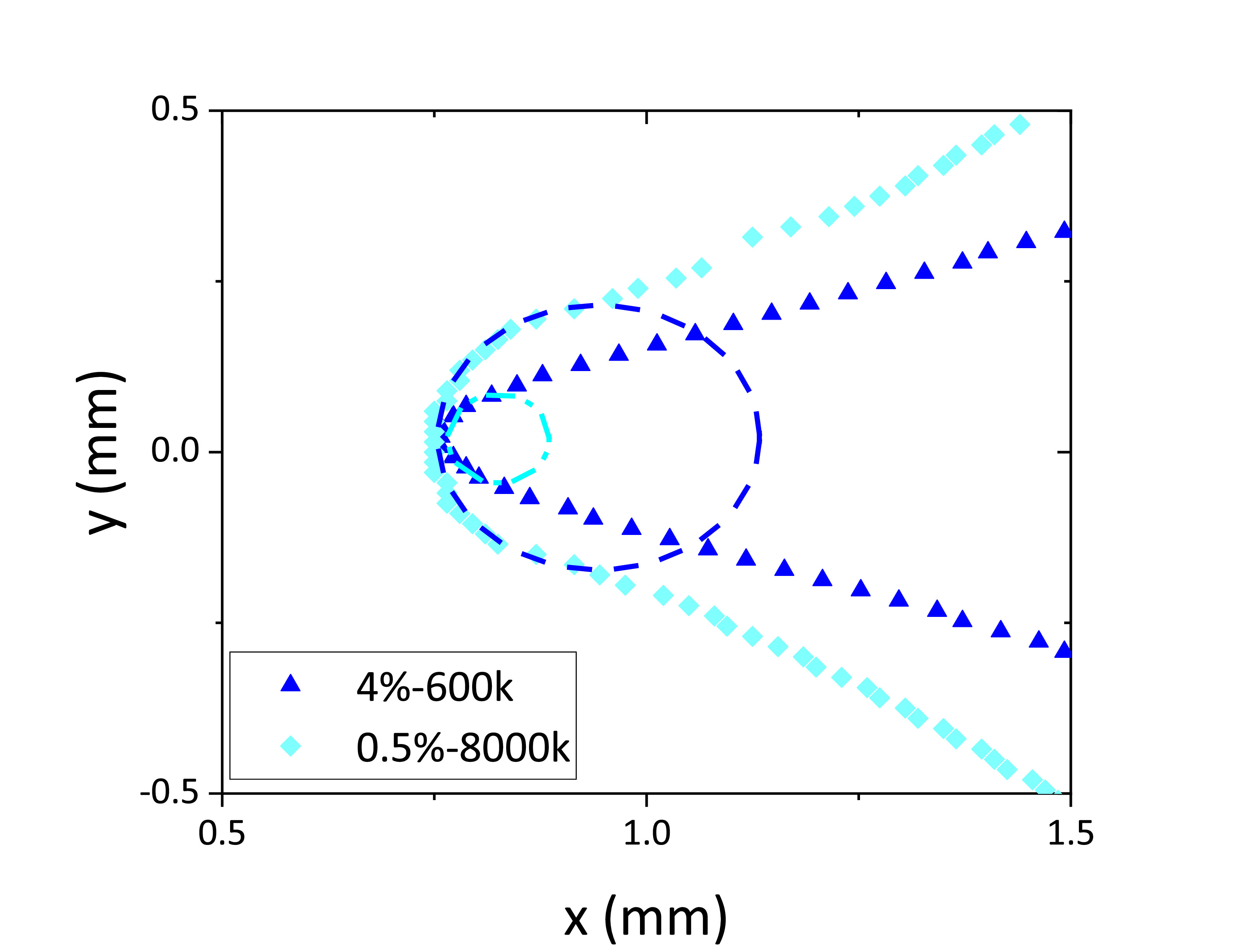}
\caption{Top: The bridge profile in the top view of drop coalescence at approximately bridge radius of \SI{0.75}{\milli \meter} for water and aqueous glycerin solution. To estimate the curvature of the neck region, a circle is fitted to this region with $r_{nk}$. 
Bottom: The bridge profile of drop coalescence at approximately bridge radius of \SI{0.75}{\milli \meter} for 4\% 600k and 0.5\% 8000k. }
\label{Profile}
\end{figure}

One of the main features of drop coalescence is the study of the evolution of the liquid bridge profile.
In recent publications on drop spreading \cite{bouillant2022rapid} and drop coalescence \cite{dekker2022elasticity} of polymer solutions, the shape evolution of the bridge is studied and compared to water. 
The authors claimed that in the case of polymer solution, due to the singular polymer stress, the liquid bridge profile becomes sharper. 
To test this idea, we extracted the radius of curvature of the contact line in the bridge from top-view images for different liquids at a fixed bridge radius (i.e. fixed deformation) of $r_b = \SI{0.75}{\milli \meter}$, Fig.~\ref{Profile}.
For Newtonian cases, as reported earlier, increasing the viscosity leads to a sharper neck and $r_{nk}$ decreases \cite{thoroddsen2005coalescence}.
Thoroddsen et al. \cite{thoroddsen2005coalescence} found a direct correlation between total curvature ($\kappa_{t}=\frac{1}{r_{nk}}-\frac{1}{r}$) and the liquid viscosity ($\kappa_{t}\sim \eta^{\frac{1}{3}}$). 
In the current work, the same behavior is observed for Newtonian cases and intermediate molar mass polymer solutions (300, 600 and \SI{1000}{(\kilo \gram \per \mol)}, see Fig.~\ref{Profile}).  
In these cases, increasing the polymer concentration and (or) polymer molar mass will result in an increase in viscosity.
At these intermediate molar masses, the polymers can contribute to viscosity more quickly than large polymers.
To illustrate this fact, we should consider the two cases in Fig.~\ref{Profile} (bottom row).
In both cases, the bridge radius is \SI{0.75}{\milli \meter} and the time after first contact is on the order of \SI{0.7}{\milli \second}. 
The zero shear rate viscosity for the sample with longer polymer chain (0.5\% 8000k) is twice that of the other sample (4\% 600k).
Based on Thoroddsen et al. \cite{thoroddsen2005coalescence}, the neck radius ($r_{nk}$) for the 8000k sample should be smaller than the other sample, but this is not the case.
The main reason is that the larger polymers have a longer relaxation time and contribute to the viscosity later. 
This observation is inline with the observation for the capillary damping timescale.
In addition, the effect of surface tension must be considered, since polymers are surface active, as shown in the previous section. 
The presence of surfactant molecules (e.g. PEO molecules) can modify the drop coalescence rate as well as the bridge profile.
The simulation showed that the presence of surfactant molecules makes the bridge profile sharper \cite{lu2012coalescence}. 
During the initial moments of coalescence, the surface area near the bridge (connecting two drops) is reduced. 
This reduction in surface area leads to an accumulation of polymer molecules near the bridge. 
This irregularity in surfactant concentration imposes a Marangoni stress and flow on the surface away from the bridge. In this sense, increasing the viscosity and (or) having surfactant affect the bridge profile in the same way.
Another point to consider is that the strength of Marangoni flow depends on the size of the surfactant, and in the current study the polymers used are larger than common surfactants. 
In conclusion, the present methods do not allow us to quantify the strength of Marangoni flow or to decouple this effect from the viscosity effect. 

\begin{figure}
   \includegraphics[width=0.75\linewidth]{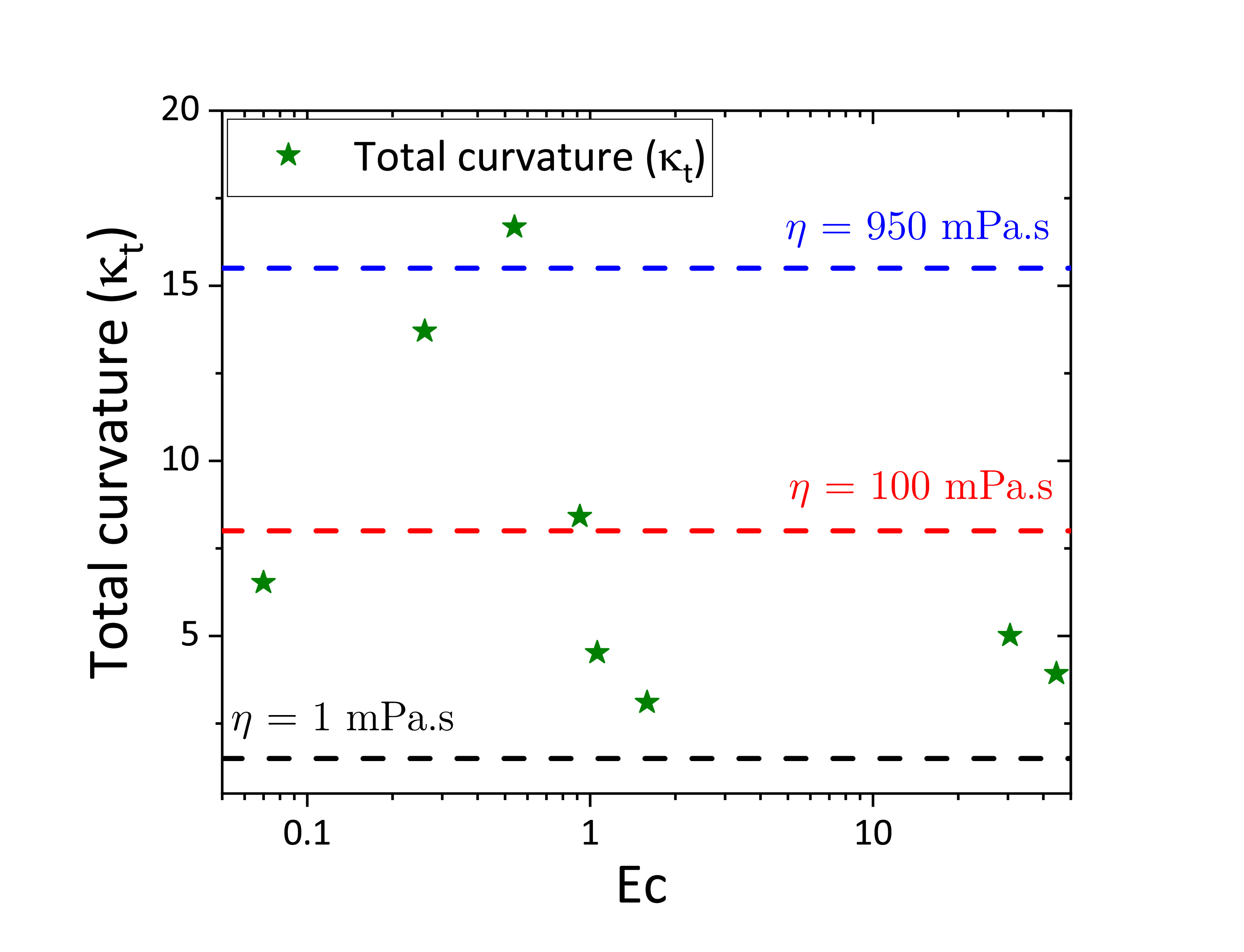}
   \includegraphics[width=0.75\linewidth]{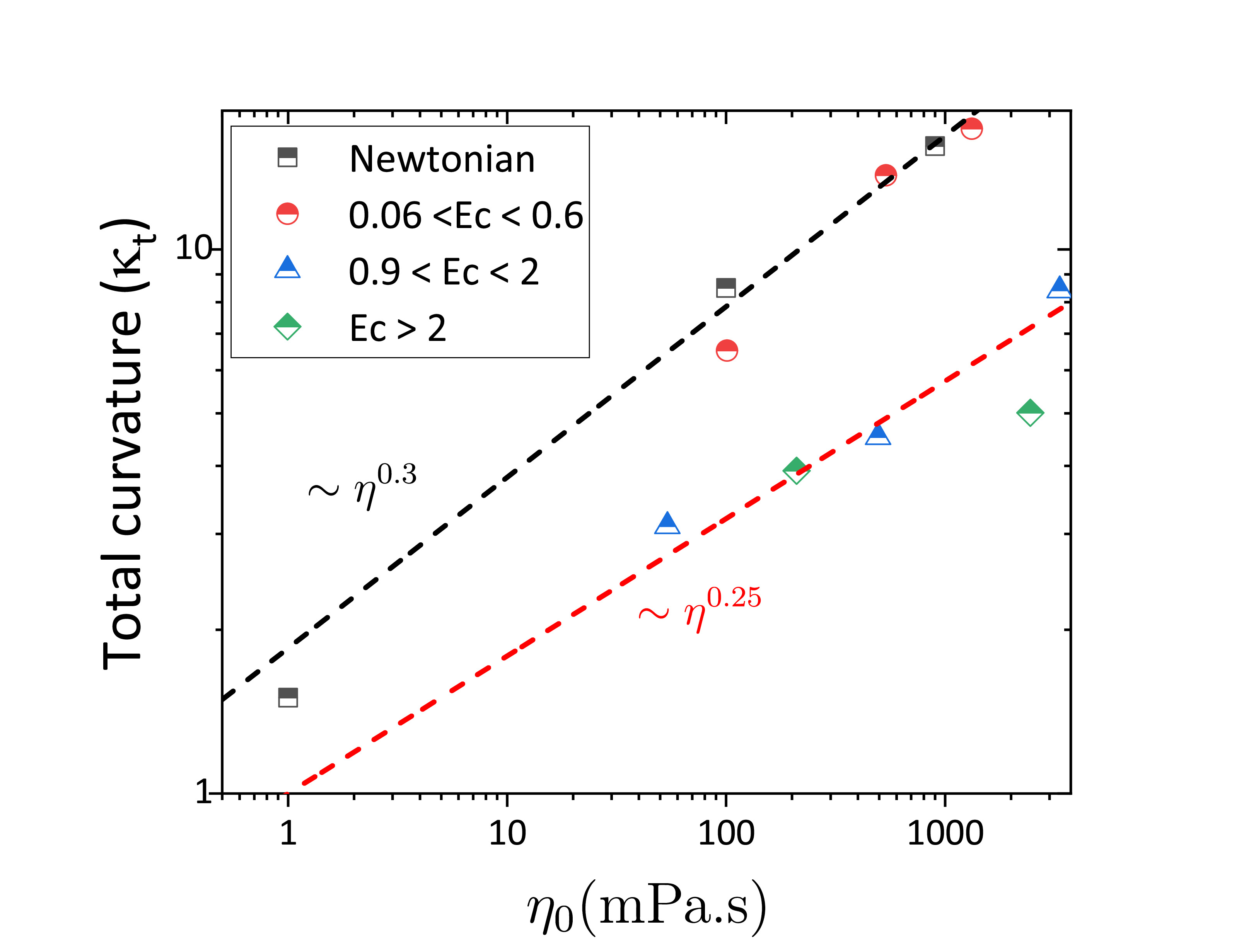}

\caption{Top: The total curvature of liquid bridge ($\kappa_{t}=\frac{1}{r_{nk}}-\frac{1}{r}$) at approximately bridge radius of \SI{0.75}{\milli \meter} for polymer solution as a function of elastocapillary number $\textup{Ec}$. 
Bottom: The total curvature of liquid bridge as a function of zero shear rate viscosity, dashed lines are guide to eyes. }
\label{totalcurvature-Ec}
\end{figure}

For a more quantitative insight, when we used the higher molar mass sample (\SI{8000}{\kilo \gram \per \mol}), the neck became wider again despite the higher zero shear rate viscosity (Fig.~\ref{Profile} bottom).
The neck radius in this case is $r_{nk}=\SI{0.19}{\milli \meter}$, which is close to the value for water ($r_{nk}=\SI{0.22}{\milli \meter}$).
This means that the Laplace pressure ($\gamma(\frac{1}{r_{nk}}-\frac{1}{r})$) inside the bridge is of the same order for both cases although the polymer solution has three order of magnitude higher zero shear rate viscosity.
This is an indication that the polymers do not contribute to the viscosity at the first moment of coalescence.
Again, the parameter that combines the viscosity, surface tension and elasticity (polymer relaxation time scale) is the elastocapillary number. 
Thus, the total curvature ($\kappa_{t}$)  at $r=\SI{0.75}{\milli \meter}$ as a function of the elastocapillary number (using the polymer relaxation time as the time scale) is plotted in Fig.~\ref{totalcurvature-Ec} top.
Here the inverse dependence on the elastocapillary number is observed. 
Around the elastocapillary number equals the unity, a maximum occurs. 
The relation between the total curvature ($\kappa_{t}$) and viscosity for Newtonian cases and low elastocapillary numbers (\textup{Ec} $< 0.6$) is similar to that is observed by another group \cite{thoroddsen2005coalescence}. 
This means that in this range, the main reason for a sharper neck is the viscosity.
The dependency on viscosity reported by Thoroddsen et al. \cite{thoroddsen2005coalescence} is $\kappa_{t}\sim \eta^{\frac{1}{3}}$ which is in a good agreement with our results. 
For higher molar masses (i.e. \textup{Ec}$\sim 1$) the dependence starts to change and the power decreases to 1/4.
For even higher molar masses (\textup{Ec}$ > 2$) the power decreases to 1/10.
However, in all cases, increasing the viscosity increases the total curvature (i.e. sharper neck). 
In conclusion, the change in spatial evolution can be explained by the viscosity effect and we cannot decouple this effect from what is reported as the singularity in polymer stress \cite{dekker2022elasticity,bouillant2022rapid}.

\section{Conclusion}

We studied the coalescence of identical polymer solution sessile drops on a hydrophobic substrate. 
We captured the coalescence process from side and top views to measure liquid bridge height and bridge radius over time.
The present study considered a wide range of polymer molar masses and concentrations. 
This wide range of fluids showed different viscoelasticities. 
The main parameter that captures this behavior is the elastocapillary number (the ratio between the polymer relaxation time scale and the viscous time scale).  
The early stage of coalescence exponents for the bridge width and height (i.e. $\alpha$ and $\beta$) decrease with increasing elastocapillary number and reach a minimum around unity. 
After this minimum, the exponents increase rapidly and reach a plateau.
For high elastocapillary numbers (i.e. high molar masses) we have shown that in the first few milliseconds the behavior looks like a quasi inviscid Newtonian case. 
The other indication of the quasi inviscid behavior at early times is the presence of a capillary wave for high elastocapillary numbers ($Ec > 2$), which contradicts the high zero shear rate viscosity. 
Raman microscopy is also used to estimate the surface activity of PEO and its concentration in the drop and over time. 
We found out that the PEO accumulates on the surface of the drop and decreases in the bulk of the drop. 
In the last section, the bridge profile is studied.
The same dependency to elastocapillary number is observed for the capillary damping timescale and the total curvature.
In summary, the results illustrate the strong coupling between non-Newtonian polymer dynamics and short-time wetting processes. 
This coupling is particularly pronounced when the internal timescale of the polymer solution coincides with the external timescale of the wetting process.

\begin{acknowledgments}
This study was funded by German Research Foundation (DFG) within the Collaborative Research Centre 1194 “Interaction of Transport and Wetting Processes,” Project- ID 265191195, subprojects A02, A06 and A07.
\end{acknowledgments}

\section*{Data Availability Statement}

The data that support the findings of this study are available from the corresponding author upon reasonable request.


\bibliography{apssamp}

\end{document}